\documentclass[usenatbib,usegraphicx]{mn2e}
\usepackage{amssymb}
\usepackage{color}
\usepackage{graphicx}

\newcommand{\halfspace}{\hspace{1pt}}

\newcommand{\MSun}{\mathop{\rm M_\odot}\nolimits}

\newcommand{\kms}{\mathop{\rm km \ s^{-1}\,}\nolimits}
\newcommand{\Lya}{Ly$\alpha$}

\newcommand\HI{{\hbox{H\halfspace$\rm \scriptstyle I$}}}

\newcommand\lsim{~\lower.5ex\hbox{$\buildrel < \over \sim$}~}
\newcommand\gsim{~\lower.5ex\hbox{$\buildrel > \over \sim$}~}

\title[Gas around galaxies:\ hydrogen absorption signatures]{Gas around
galaxy haloes - II: hydrogen absorption signatures from the
environments of galaxies at redshifts $2<z<3$}
       
\author[Avery Meiksin, James S. Bolton, Eric R. Tittley]{
        Avery Meiksin$^{1}$\thanks{E-mail:\ A.Meiksin@ed.ac.uk (AM)},
        James S. Bolton$^{2}$, Eric R. Tittley$^{1}$\\
        $^{1}$SUPA\thanks{Scottish Universities Physics Alliance},
	Institute for Astronomy, University of Edinburgh,
        Blackford Hill, Edinburgh\ EH9\ 3HJ, UK\\
        $^{2}$School of Physics and Astronomy, University of Nottingham,
        University Park, Nottingham\ NG7\ 2RD, UK}

\shortcites{}

\begin{document}

\date{Accepted . Received ; in original form }
\pagerange{\pageref{firstpage}--\pageref{lastpage}} \pubyear{2015}
\maketitle
\label{firstpage}

\begin{abstract}

We compare predictions of large-scale cosmological hydrodynamical
simulations for neutral hydrogen absorption signatures in the vicinity
of $10^{11}$--$10^{12.5}\,M_{\odot}$ haloes with observational
measurements.  Two different hydrodynamical techniques and a variety
of prescriptions for gas removal in high density regions are examined.
Star formation and wind feedback play only secondary roles in the
\HI\ absorption signatures outside the virial radius, but play
important roles within. Accordingly, we identify three distinct
gaseous regions around a halo:\ the virialized region, the
mesogalactic medium outside the virial radius arising from the
extended haloes of galaxies out to about two turnaround radii, and the
intergalactic medium beyond.  Predictions for the amount of absorption
from the mesogalactic and intergalactic media are robust across
different methodologies, and the predictions agree with the amount of
absorption observed around star-forming galaxies and QSO host
galaxies. Recovering the measured amount of absorption within the
virialized region, however, requires either a higher dynamic range in
the simulations, additional physics, or both.

\end{abstract}


\section{Introduction}
\label{sec:Intro}

The environment of a forming galaxy is expected to be violent and
complex. Cooling and accreting gas will be thermally unstable on
multiple scales, leading to star formation.  Supernovae-driven winds
from young massive stars will drive back accreting gas. An active
galactic nucleus (AGN), particularly in more massive galaxies, will
produce jets that will further drive out gas, while the accompanying
compression of gas latitudinal to the jet may enhance star
formation. The evolution will be occasionally punctuated by major
mergers with other haloes.

Cosmological simulations have begun filling in the physical details of
these processes. A picture is emerging of two modes of gas accretion,
a ``hot mode'' in galaxies sufficiently massive for most gas to shock
heat to virial temperatures before the onset of gradual cooling, and a
``cold mode'' in galaxies in which much of the gas does not sustain
high post-shock temperatures and instead flows in primarily as cool
gas streams \citep{2000MNRAS.316..374K, 2005MNRAS.363....2K,
  2006MNRAS.368....2D, 2011MNRAS.415.2782V}, a scenario supported by
theoretical arguments \citep{2003MNRAS.345..349B}.

The cooling gas feeds star formation in the galaxies. Simulations,
however, tend to over-predict the amount of stars formed. Supernovae
have long been suspected of suppressing star formation in galaxies,
from ellipticals \citep{1974MNRAS.169..229L, 1986ApJ...307..415S} to
dwarfs \citep{1986ApJ...303...39D}, to bring the amount of stars
formed into agreement with observations. This view has been backed by
large-scale hydrodynamical simulations, strongly suggesting that
supernovae-driven wind feedback may regulate the star formation rate
in galaxies \citep[e.g.][]{2002MNRAS.330..113K, 2010MNRAS.402.1536S,
  2013MNRAS.435.2931H}. AGN jets may furthermore regulate star
formation in the more massive galaxies
\citep[e.g.][]{2011MNRAS.415.2782V, 2013MNRAS.435.2931H}.

However, direct observational evidence for these mechanisms is
scanty. While hydrogen and metal emission lines show evidence for
winds in star-forming galaxies at redshifts $z>1$ with velocities
exceeding the escape velocities of the haloes
\citep{2011ApJ...733..101G}, the extent of the winds and their impact
on the surrounding gas are still unknown. Maps of gas emission in
active galaxies reveal winds extending over tens of kiloparsecs, but
so far only at low redshifts \citep{2014MNRAS.441.3306H}.
Consequently, the search for cold streams in absorption in the spectra
of background Quasi-stellar Objects (QSOs) or bright galaxies has
attracted increasing recent attention. The availability of large QSO
catalogs with a high sky density \citep[e.g.][]{2009ApJS..182..543A,
  2012ApJS..203...21A}, has presented an opportunity to study the
gaseous environments of star-forming galaxies (SFGs) and AGNs through
absorption measurements with unprecedented detail. Observations of
intergalactic \HI\ show a rise in the \HI\ optical depth near
Lyman-break galaxies (LBGs) \citep{2010ApJ...717..289S,
  2011MNRAS.414...28C, 2012ApJ...751...94R} and QSOs
\citep{2013ApJ...776..136P}, and thus evidence for extended cool gas
surrounding galactic haloes.

Several galaxy formation simulations have sought to establish
consistency with these data.  Of particular focus have been absorbers
optically thick at the photoelectric edge, Lyman limit systems (LLS)
and damped Ly$\alpha$ absorbers (DLAs). Simulations suggest a large
fraction of these absorption systems at $z>2$ may be accounted for by
cold-mode accretion of circumgalactic gas \citep{2011MNRAS.412L.118F,
  2011MNRAS.418.1796F, 2012MNRAS.421.2809V, 2013ApJ...765...89S}.
However, numerical models have been less successful at reproducing the
observed absorption around QSOs relative to SFGs \citep[but see recent
  work by][]{2015arXiv150305553R}. While the covering fraction of
optically thick systems of SFGs in simulations matches the
observations within the errors, most previous simulation predictions
fall short of the covering fractions measured around QSOs
\citep{2014ApJ...780...74F, 2014arXiv1409.1919F}. Agreement with the
trend of increasing \HI\ absorption with decreasing impact parameter
is also reported for the SFG sample of \citet{2010ApJ...717..289S}
\citep{2011MNRAS.418.1796F, 2012MNRAS.424.2292G, 2013ApJ...765...89S},
but for a sample of QSOs, simulations greatly underpredict the
observed amount of absorption by circumgalactic gas
\citep{2013ApJ...776..136P}.

In this paper, we examine the \HI\ absorption around galaxy haloes in
large scale cosmological hydrodynamical simulations, by considering
both the circumgalactic gas and the extended region surrounding it as
it merges into the intergalactic medium (IGM).  Previous simulations
of the circumgalactic medium (CGM), generally defined as gas within
300~kpc (proper) of a galaxy, have instead typically been zoom-in
models focussed on the immediate surroundings of a galaxy, extending
out to at most a few virial radii. They have not therefore been able
to fully exploit the growing amount of data probing the gas around
galaxies to distances of several comoving megaparsecs.  Exceptions are
\citet{2013MNRAS.433.3103R}, who used the median Ly$\alpha$ optical
depth measurements of \citet{2012ApJ...751...94R} to constrain the
masses of the SFG haloes in the Keck Baryonic Structure Survey, and a
very recent study by \citet{2015arXiv150305553R} using the EAGLE
simulation \citep{Schaye2015}.

The analysis presented here is instead based on the {\texttt{Enzo}}
and {\texttt{GADGET-3}} simulations described in
\citet{2014MNRAS.445.2462M} (hereafter Paper I). These are of
sufficiently high resolution and volume to recover the hydrogen
absorption properties of the IGM.  In Paper I we demonstrated that
these IGM simulations make consistent predictions for the
intergalactic gas properties beyond the halo turn-around radii,
despite their differing numerical algorithms.  However, the gas
properties on smaller scales are highly dependent on star formation
and feedback implementations \citep[see also][]{2015MNRAS.448..895S,
  2015arXiv150302665N}.  We shall use these simulations to construct
spectra to examine how well the models match the observed statistics
of \HI\ absorption around galaxies and QSOs.  We differ from previous
work, however, by first taking a step back and asking whether the
absorption signatures of gas around galaxies show any evidence for
star formation even without feedback. We then seek to establish the
physical extent of excess hydrogen absorption over the contribution
from the diffuse IGM around galaxies. We finally ask what
observational evidence hydrogen absorption signatures provide for wind
feedback in addition to star formation.

This paper is organized as follows. In Sec.~\ref{sec:sims} we
summarise the numerical simulations used in this work. The resulting
\HI\ optical depths and covering fractions around haloes are presented
in Sec.~\ref{sec:results}, and these results are discussed in detail
in Sec.~\ref{sec:discussion}. We present our conclusions in
Sec.~\ref{sec:conclusions}, and simulation convergence tests are
discussed in an Appendix.  All results are for a flat $\Lambda$CDM
universe with the cosmological parameters $\Omega_m=0.28$,
$\Omega_bh^2=0.0225$ and $h=H_0/100~\kms\,{\rm Mpc}^{-1}=0.70$,
representing the present-day total mass density, baryon density and
Hubble constant, respectively. The power spectrum has spectral index
$n_{\rm s}=0.96$, and is normalized to $\sigma_{8}=0.82$, consistent
with the 9-year {\it Wilkinson Microwave Anisotropy Probe} ({\it
  WMAP}) data \citep{2013ApJS..208...19H}.

\section{Numerical simulations}
\label{sec:sims}
\subsection{Cosmological hydrodynamics codes}

\begin{table*}
  \centering  
    \begin{minipage}{180mm}
      \begin{center}
        \caption{Summary of the simulations performed in this work.
          The columns, from left to right, list the simulation name,
          the box size in comoving Mpc, the number of resolution
          elements in the simulation, the code used for the run, the
          star formation prescription and whether or not the model
          includes supernovae-driven winds. }
    \begin{tabular}{l c c c c r}
      \hline\hline  
      Name       & Box size     & Resolution & Method    & Star  
      formation & Winds       \\
                 & [Mpc] & elements \\
      \hline  
      G30qLy$\alpha$  &      30  &  2$\times$512$^3$ &  {\texttt{GADGET-3}}  &
      qLy$\alpha$ & N \\
      G30sfnw         &      30  &  2$\times$512$^3$ &  {\texttt{GADGET-3}}  &
      SH03 & N \\
      G30sfw          &      30  &  2$\times$512$^3$ &  {\texttt{GADGET-3}}  &
      SH03 & Y \\
      \hline  
      E30\_512  &           30  &  512$^3$ &  {\texttt{ENZO}}  &
      none & N \\
      E60\_1024 &           60  & 1024$^3$ &  {\texttt{ENZO}}  &
      none & N \\
      \hline  
    \end{tabular}
    \label{tab:sims}
\end{center}
  \end{minipage}
\end{table*}
 
The numerical simulations are performed using \texttt{ENZO}, v.2
\citep{2014ApJS..211...19B}, and \texttt{GADGET-3}, an updated version
of the publicly available code \texttt{GADGET-2} \citep[last described
  by][]{2005MNRAS.364.1105S}.  The runs are summarized in
Table~\ref{tab:sims}, and further details may also be found in Paper
I.  As we focus on high redshift haloes in this work, all the runs
were stopped at $z=2$. The Eulerian \texttt{ENZO} simulations were
performed on the top-level grid only.  One simulation uses $512^3$
mesh zones and $512^3$ cold dark matter particles in a box 30~Mpc
(comoving) on a side (E30\_512), and a second has $1024^3$ mesh zones
and $1024^3$ cold dark matter particles in a 60~Mpc box (E60\_1024).
The grid zones have a comoving width of 58.6~kpc. Since
{\texttt{ENZO}} uses the particle-mesh method, the force resolution is
limited to two grid zones, adequate for resolving the Jeans length of
the photoionized gas. The smoothed particle hydrodynamics simulations
performed with {\texttt{GADGET-3}} were run with $512^3$ cold dark
matter particles and $512^3$ gas particles in a box $30$~Mpc
(comoving) along a side, with a force softening scale of 1.4~kpc
(comoving). The baryonic and dark matter particle masses are
$1.3\times10^6\,M_\odot$ and $6.4\times10^6\,M_\odot$, respectively.

No gas is removed from the \texttt{ENZO} simulations, producing a ``no
star formation'' control sample. However, because of the high
densities and temperatures reached in the Lagrangian \texttt{GADGET-3}
simulations, gas must be removed in rapidly cooling regions. We adopt
the star formation prescription of \citet{2003MNRAS.339..289S} (SH03)
in one model (G30sfnw), and include their supernovae feedback
prescription with a wind velocity of $484\kms$ in another (G30sfw).
Note, however, that any conclusions drawn from its use are exclusive
to this particular algorithm \citep[see also
  e.g.][]{2008MNRAS.387.1431D}.  Indeed, absorption measurements like
those we examine may provide a basis for discriminating between
different feedback models.  A comparison simulation, discussed in the
Appendix, is also performed using the ``quick Ly$\alpha$'' method
(G30qLy$\alpha$), for which all gas cooler than $10^5$~K and with an
overdensity exceeding 1000 is converted into collisionless ``star''
particles (without feedback).  This is a computational trick used in
IGM simulations to significantly speed up the computation
\citep{2004MNRAS.354..684V}.  We use it as an alternative for testing
the sensitivity of the results on the gas removal scheme.

Both the {\texttt{ENZO}} and {\texttt{GADGET-3}} codes solve the
time-dependent ionization equations for hydrogen and
helium. Photoionization, collisional ionization and radiative and
dielectric recombinations are included. Thermal balance includes
cooling by radiative and dielectronic recombination, collisional
ionization and excitation and thermal bremsstrahlung losses as well as
Compton cooling off the cosmic microwave background. The codes adopt
the photoionization and photoheating rates of
\citet{2012ApJ...746..125H}, except we adjusted the photoionization
heating rate of singly ionized helium to reproduce the evolution of
the IGM temperature in \citet{2011MNRAS.410.1096B} for
$\gamma=1.3$. The codes used identical atomic rates for the heating
and cooling, as summarized in \citet{2009RvMP...81.1405M}, except for
adopting the \HI\ electron excitation and collisional cooling rate of
\citet{1991ApJ...380..302S}.  We have not included any photoionization
from the central galaxies or QSOs, which could reduce the amount of
absorption from high column density systems
\citep{MiraldaEscude2005,Schaye2006}. However, to date there is no
firm evidence for a widespread transverse proximity effect from either
SFGs or QSOs \citep{2004ApJ...610..642C, 2008MNRAS.391.1457K}.  We
also do not include the self-shielding of \HI\ absorbers in our main
analysis, although in the Appendix we show this has only a small
effect on the integrated absorption statistics examined here.

\subsection{Extraction of \Lya\ absorption statistics}

The \Lya\ absorption spectra are computed at restframe velocity $v$
along a line of sight through a periodic box of comoving side $L$ as
$\exp[-\tau_\alpha(v)]$, where
\begin{equation}
  \tau_\alpha(v) =
  \frac{\sigma_\alpha\lambda_\alpha}{\pi^{1/2}}\frac{L}{1+z} \int dx
  n_{\rm HI}(x) b(x)^{-1} e^{-(v-v_{\rm los})^2/b^2}.
\label{eq:taua}
\end{equation}
Here $v_{\rm los}=xH(z)L/ (1+z)+v_{\rm pec}$ at dimensionless distance
$x$, in box side units, for Hubble parameter $H(z)$ at redshift $z$
and line-of-sight peculiar velocity $v_{\rm pec}$.  The Doppler
parameter $b(x)=[2k_B T(x)/m_{\rm H}]^{1/2}$ for gas at temperature
$T(x)$, $\sigma_\alpha=(\pi e^2/ m_ec)f_\alpha$ is the total
cross-section for Ly$\alpha$ scattering with upward oscillator
strength $f_\alpha\simeq0.4162$, $\lambda_\alpha\simeq1215.67\rm\,\AA$
is the Ly$\alpha$ wavelength, $k_B$ is the Boltzmann constant and
$m_{\rm H}$ is the mass of a hydrogen atom. For the \texttt{ENZO}
simulations, the gridded \HI\ density, temperature and peculiar
velocity fields are used. For the \texttt{GADGET-3} simulations, the
spectra are computed along rays using \HI-weighted contributions from
the particles \citep{1996ApJ...457L..51H}. To match the resolution of
the observations, we smooth the resulting spectra with a Gaussian of
FWHM $8\kms$ for comparison with \citet{2012ApJ...751...94R} and FWHM
$125\kms$ for comparison with \citet{2013ApJ...776..136P}.

We extract spectra for a range of impact parameters around haloes
identified in the simulations using the grid-based halo finding
algorithm described in Paper I.  We consider three spectral signatures
and their dependence on impact parameter: the 2D absorption as
measured by the median Ly$\alpha$ optical depth around the haloes, the
Ly$\alpha$ absorption equivalent width and the fractional deviation of
the Ly$\alpha$ absorption from the mean intergalactic value.  Note the
latter two statistics are integrated quantities.  The first of these
is the equivalent width around a halo at measured velocity $v_{\rm
  halo}$ along a line of sight to a background QSO, which is computed
as
\begin{equation}
  w(b_\perp, \Delta v) = \frac{\lambda_\alpha}{c}\int_{v_{\rm halo}-\Delta v/2}^{v_{\rm halo}+\Delta v/2}\,dv \left[1-e^{-\tau_\alpha(b_\perp, v)}\right],
\label{eq:ewDv}
\end{equation} 
over a velocity window of width $\Delta v$ centred on the position of
the halo, displaced transversely by an amount $b_\perp$. Following
\citet{2012ApJ...751...94R}, we also consider an alternative
determination of the equivalent width by dividing the transmission by
a factor meant to correct for the errors incurred in continuum fitting
low-resolution spectra. The second integrated absorption statistic we
consider is the fractional deviation, $\delta_F$, of the absorption
from the mean intergalactic value, introduced by
\citet{2013ApJ...776..136P}. This quantity references the equivalent
width to the baseline value expected from the IGM,
\begin{equation}
\delta_F(b_\perp, \Delta v)=\frac{w(b_\perp, \Delta v) - w_{\rm
    IGM}}{\Delta\lambda - w_{\rm IGM}}.
\label{eq:dFDv}
\end{equation}
Here $\Delta\lambda = \lambda_\alpha\Delta v/c$ and $w_{\rm
  IGM}=\Delta\lambda[1-\exp(-\tau_{\rm eff})]$, where $\tau_{\rm eff}$
is the effective optical depth of the IGM. We adopt the values from
\citet{2013MNRAS.430.2067B} for $\tau_{\rm eff}$. Our simulations
recover the mean transmission according to these values to better than
2--4 per cent at $2.0 < z < 2.4$, and to better than 1 per cent at
$2.4<z<2.8$. Rather than renormalizing the photo-ionization rates in
each simulation to recover the values of $\tau_{\rm eff}$ exactly, we
take these small deviations to be the uncertainty in the model
predictions of the amount of IGM absorption. All the error bars shown
for the model predictions are errors in the mean. The error in the
mean IGM effective optical depth may be added linearly to the error
bars as a systematic uncertainty.

Lastly, in the Appendix we present a series of convergence and
parameter tests on these statistical measures.  In summary, we find:\
(a)\ a box size of 60~Mpc is preferred over 30~Mpc for \texttt{ENZO}
simulations to capture statistics on haloes with $M_h>10^{12}\MSun$;
(b) at the resolution of the simulations, radiative transfer effects
negligibly affect the predicted integrated \HI\ absorption statistics;
(c) when excluding winds, the \texttt{GADGET-3} results for $\delta_F$
are little changed if the quick Ly$\alpha$ method is used to convert
gas to stars rather than the algorithm of \citet{2003MNRAS.339..289S};
(d) for uncertainties $\sigma_H$ in galaxy systemic velocities and a
velocity window $\Delta v$, $\delta_F$ is insensitive to $\sigma_H$
for values $\sigma_H<\Delta v/ 2$; and (e) for
$\sigma_H\approx500\kms$, a velocity window $\Delta v=2000\kms$
provides a good compromise between a sensitive dependence of
$\delta_F$ on impact parameter and small variance in its values.

\section{Results}
\label{sec:results}
\subsection{An illustrative example of \Lya\ absorption around a halo}

\begin{figure*}
\scalebox{0.55}{\includegraphics{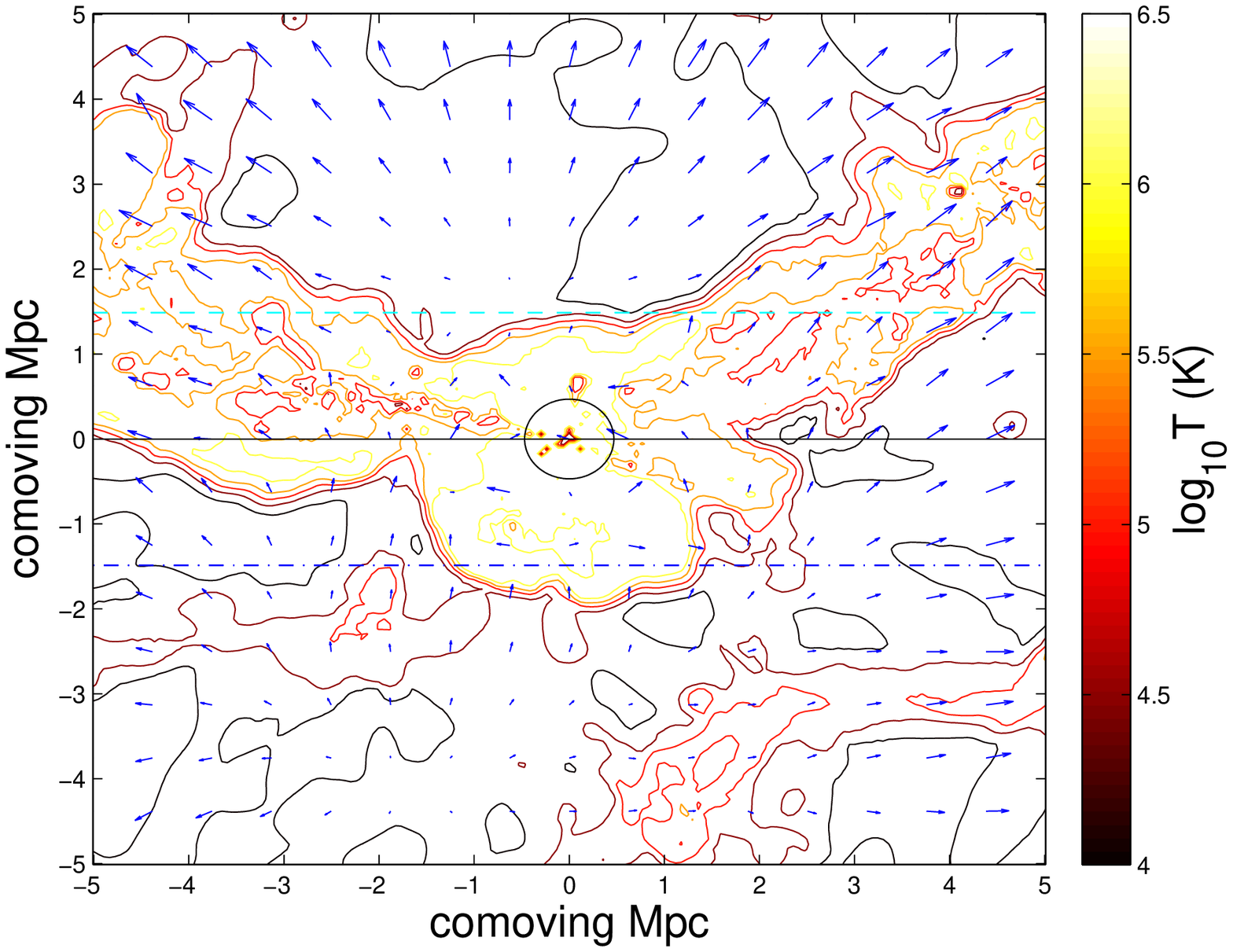}}\\
\scalebox{0.5}{\includegraphics{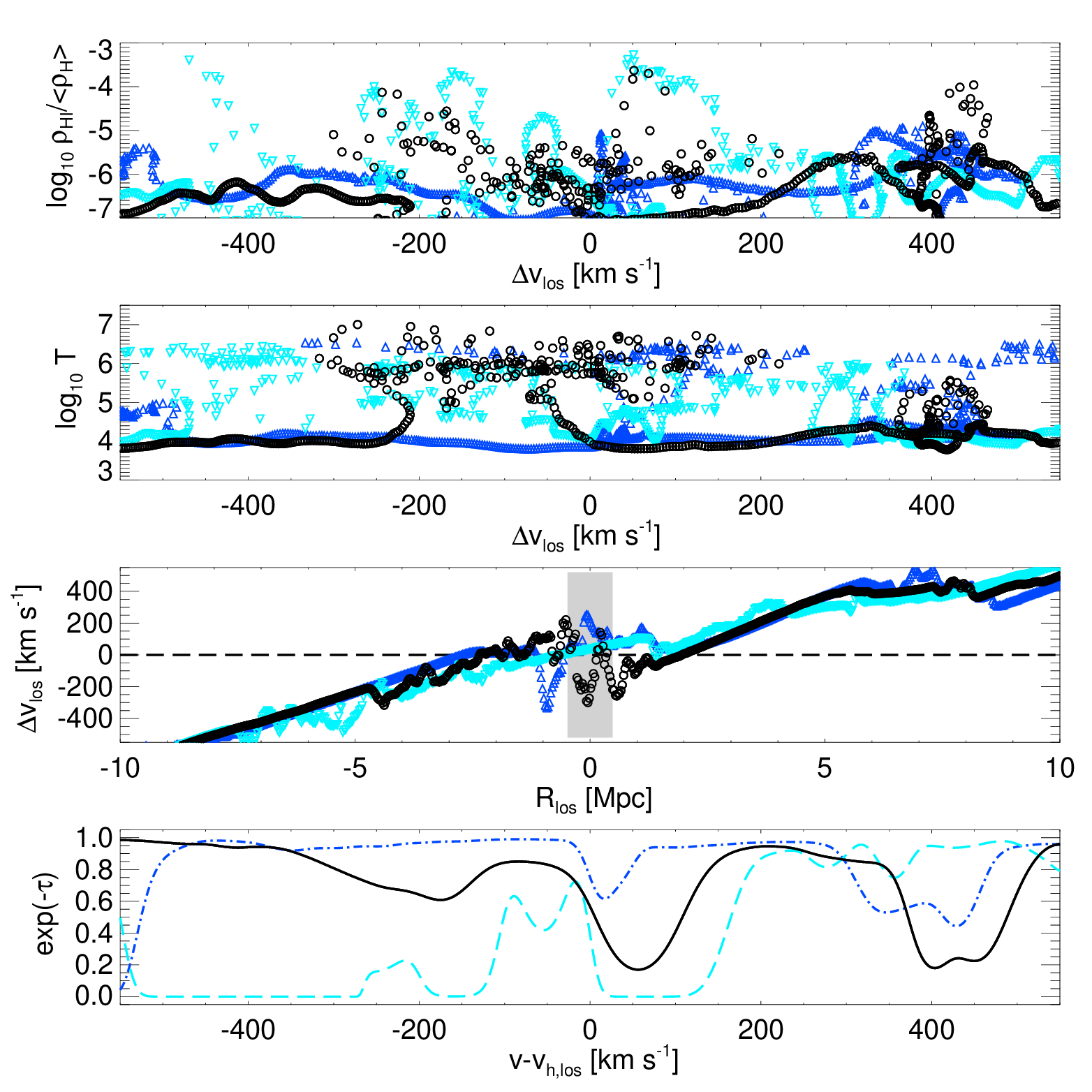}}
\vspace{-0.2cm}
\caption{{\it Top panel:} \HI-weighted temperature and proper velocity
  fields around a $3\times10^{12}\,M_\odot$ halo at $z=2$ from the
  \texttt{GADGET-3} 30~Mpc box simulation without winds (G30sfnw). The
  velocities are shown relative to the centre-of-mass of the halo,
  where an arrow length equivalent to the spacing of the axis tic
  marks corresponds to $1000\kms$. The halo virial radius is
  represented by a circle.  The three \Lya\ absorption spectra shown
  in the lower panels are drawn along the lines running parallel to
  the x-axis. {\it Lower panels:} line-of-sight plots showing the
  \HI\ fraction, $\rho_{\rm HI}/\langle \rho_{\rm H} \rangle$, and gas
  temperature $T$, as functions of the line-of-sight velocity offset
  $\Delta v_{\rm los}$ from the halo centre-of-mass at $v_{h, {\rm
      los}}$. In the second panel from the bottom, $\Delta v_{\rm
    los}$ is shown as a function of the comoving distance, $R_{\rm
    los}$, from the halo centre-of-mass.  The shaded region indicates
  the size of the virialized zone. Note that regions at multiple
  values of $R_{\rm los}$ have $\Delta v_{\rm los}\simeq0$, such that
  gas outside the hot central region contributes to absorption near
  and across the systemic velocity of the halo. The bottom panel shows
  the corresponding spectra. The open circles and solid black lines
  are for a line of sight passing through the halo centre-of-mass. The
  inverted cyan triangles (upright blue triangles) and dashed cyan
  (dot-dashed blue) lines are for a line of sight laterally offset
  from the halo centre-of-mass by $+1.5$ ( $-1.5$) comoving Mpc.}
\label{fig:halo_gadget_z2_los}
\end{figure*}

\begin{figure*}
\scalebox{0.5}{\includegraphics{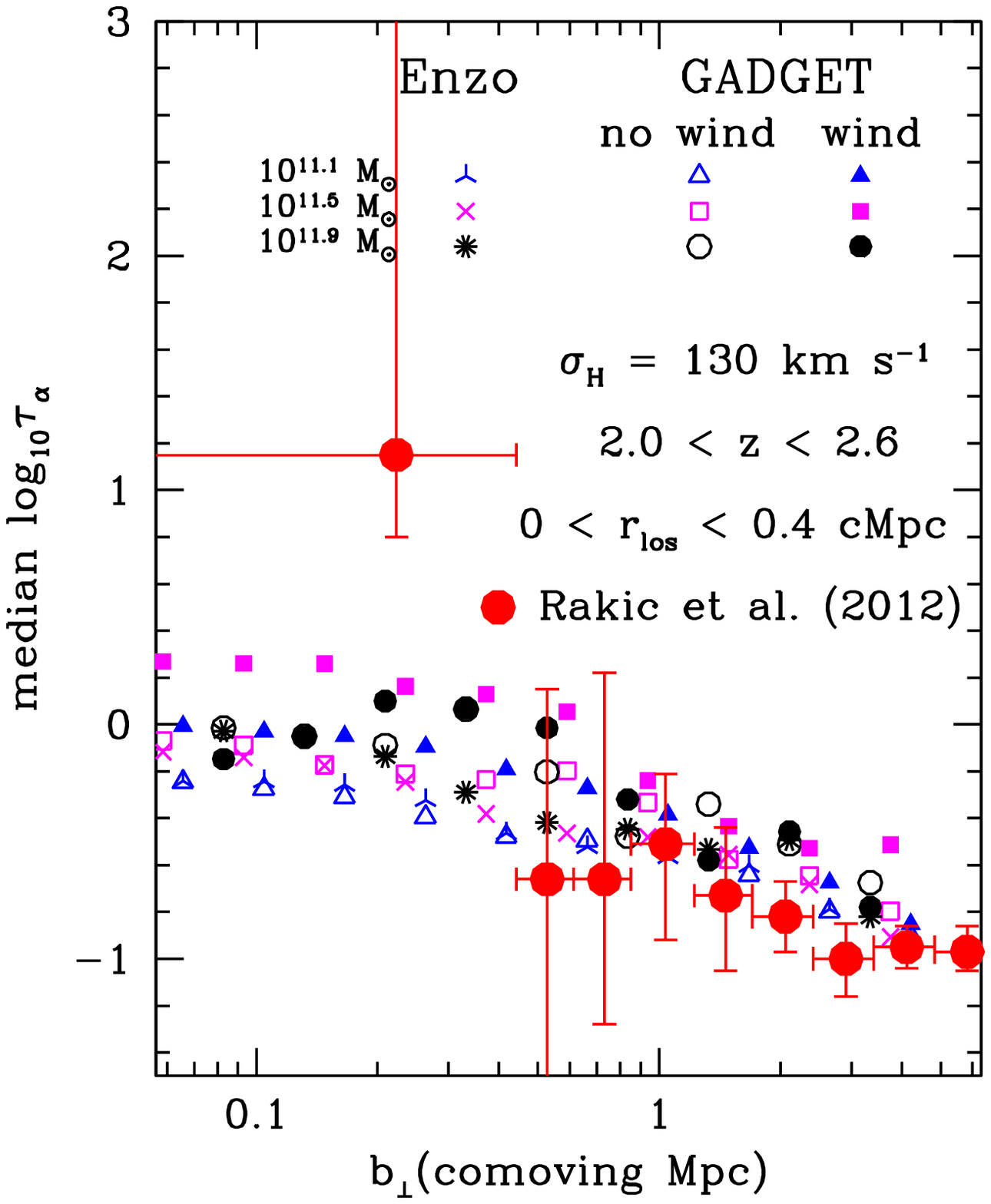}\includegraphics{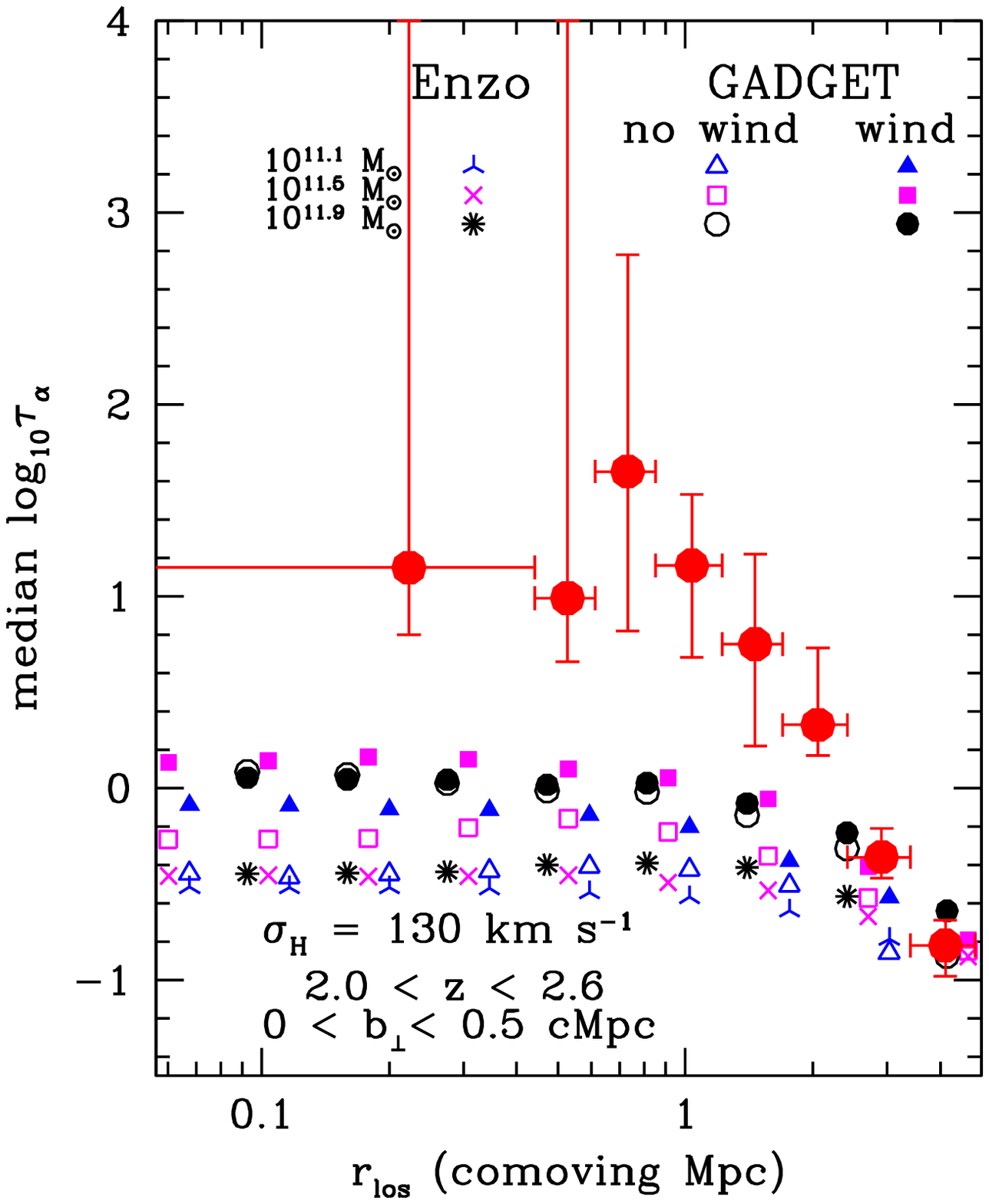}}
\vspace{-1.5cm}
\caption{{\it Left panel:} median Ly$\alpha$ optical depth, averaged
  over line-of-sight velocities relative to the halo systemic velocity
  corresponding to Hubble distances of up to 0.4 Mpc (comoving),
  against projected separation from the halo. The results are averaged
  over all haloes in a 0.2~dex wide mass bin centred at the indicated
  mass. The model predictions agree with the data except within the
  typical virial radius, $r_v\simeq 150$--$330\rm\,kpc$
  (comoving). {\it Right panel:} median Ly$\alpha$ optical depth,
  averaged over projected separations up to 0.5 Mpc (comoving),
  against line-of-sight Hubble displacement from the halo. Except for
  the regions along the spectra corresponding to Hubble distances
  beyond $\pm3$~Mpc (comoving), the models underpredict the measured
  absorption. In both panels, halo velocities include a random
  component drawn from a Gaussian distribution with standard deviation
  $\sigma_{\rm H}=130\kms$. The data points (large filled circles;
  red) are from Rakic et al. (2012). Note that for clarity of
  presentation, here and for other figures in this paper, the points
  in each mass bin are grouped with slight offsets using cubic spline
  interpolation.  }
\label{fig:taumed_comp_enzo_gadget}
\end{figure*}

It is instructive to first consider the \Lya\ absorption around an
individual halo in the cosmological simulations.  The gas velocity
field near a galactic halo is in general complex, involving Hubble
flow on large scales, and on smaller, cosmological infall, shocks and
streams of merging material. It is important to include a sufficiently
large volume around a halo to capture these effects on the simulated
spectra. Since the spectra are measured in velocity space, large
peculiar motions will furthermore scramble spatial and velocity
information.  Particularly large displacements in velocity space may
result from the cosmological infall around massive haloes.

We illustrate these effects for a $3\times10^{12}\,{\rm M_\odot}$ halo
at $z=2$ from the 30~Mpc \texttt{GADGET-3} simulation without a wind
(G30sfnw), shown in Fig.~\ref{fig:halo_gadget_z2_los}. Three
representative lines of sight are shown in the lower panels, one
through the halo centre-of-mass and two offset transversely by
$\pm1.5$ comoving Mpc. Despite the high halo gas temperatures,
reaching $10^7$~K, all three spectra show absorption within $\Delta
v\approx\pm500\kms$ of the halo centre-of-mass. This absorption arises
in part as a result of the complex velocity field of the gas.  The
panel second from the bottom in particular shows the wide range of
positions that give rise to line-of-sight velocities matching the
systemic velocity of the halo. These lines of sight include the region
within the turnaround radius of the gas, $r<r_{\rm t.a.}\simeq 6r_v$,
which is sensitive to the star formation and feedback implementation
(Paper I); the virial radius of the halo in
Fig.~\ref{fig:halo_gadget_z2_los} is $r_v=0.48$ comoving Mpc.

Examining individual absorption features, it is apparent that gas
extending out to the turnaround radius contributes to the broad
feature at $-50\kms < v-v_{\rm h, los}<150\kms$ along the line of
sight passing directly through the halo centre. Gas within the
turnaround radius also produces the weaker feature at
$v-v_{\rm h, los}\simeq25\kms$ along the line of sight displaced
laterally by -1.5~Mpc. The line of sight displaced by $+1.5$~Mpc
instead probes part of the cosmic web. The very broad absorption
feature covering $-550\kms < v-v_{\rm h, los} < -250\kms$ originates
in the broad filament attached to the top left of the halo, extending
over $-8.5\,{\rm Mpc}<R_{\rm los}<-3.9\,{\rm Mpc}$, as shown in the
panel second from the bottom.  It is dominated by a complex region
30~kpc across (proper) with a velocity width of about $160\kms$.  The
resulting absorption feature has an equivalent width of 1.2~{\rm \AA}
and an \HI\ column density of
$N_{\rm HI}=8\times10^{17}\,{\rm cm}^{-2}$, forming a Lyman limit
system along this line of sight \citep[see
also][]{2011MNRAS.412L.118F, 2011MNRAS.418.1796F, 2012MNRAS.421.2809V,
  2013ApJ...765...89S}. (Allowing for radiative transfer effects, as
discussed in the Appendix, increases $N_{\rm HI}$ to
$4\times10^{18}\,{\rm cm}^{-2}$, but has little effect on the spectrum
because the feature is saturated.)

\subsection{Median Ly$\alpha$\ optical depth}

We now turn to consider the observational signature of the neutral
hydrogen around galaxy haloes.  The most direct measurement of
Ly$\alpha$\ absorption from the gas around galaxies is the optical
depth $\tau_\alpha$. Because of its large variance, the median
optical depth is a more stable statistic than the
average. Fig.~\ref{fig:taumed_comp_enzo_gadget} shows the median
optical depths, as a function of separation from the halo centres,
averaged over simulated spectra in the redshift range $2.0<z<2.6$,
corresponding to most of the foreground galaxy redshifts in the sample
of \citet{2012ApJ...751...94R}. The left panel averages over
line-of-sight Hubble flow distances, defined as $r_{\rm los}=v_{\rm
  los}/H(z)$, within $\pm0.4$ comoving Mpc of the halo centres. The
right panel averages over impact parameters within $\pm0.5$ comoving
Mpc of the halo centres. Both panels show a general rise toward the
halo centres, although with differences in the trends. In assigning
halo velocities, a random Gaussian error with standard deviation of
$\sigma_{\rm H}=130\kms$ is included to match the data
\citep{2012ApJ...751...94R}.

\begin{figure}

\scalebox{0.5}{\includegraphics{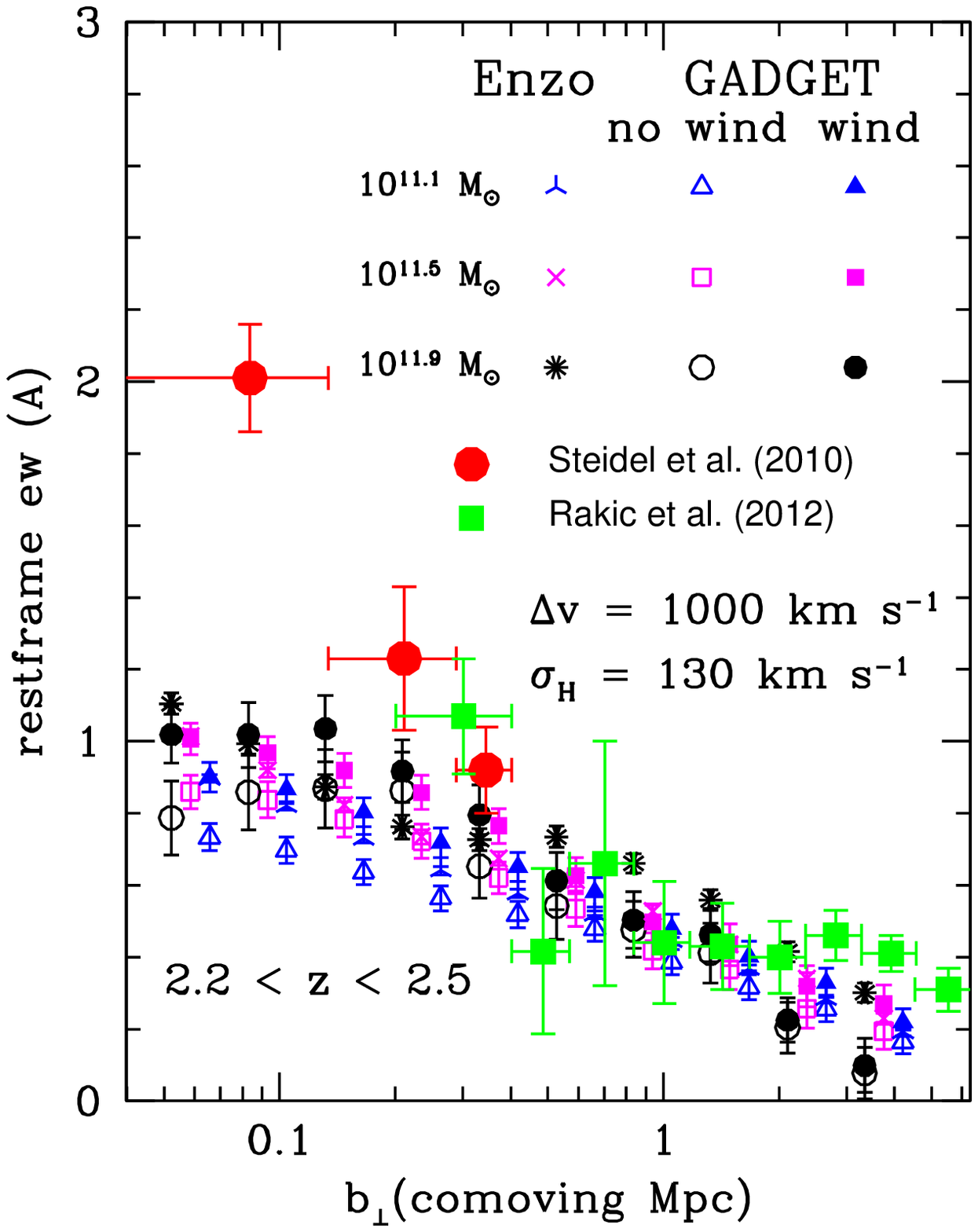}}
\vspace{-1.5cm}
\caption{Rest-frame equivalent width ($\rm \AA$) within a velocity
  window $\Delta v=1000\kms$ centred on the halo centre-of-mass
  velocity as function of line-of-sight impact parameter $b_\perp$ for
  halo masses $\log_{10}M_h=11.1$ (triangles; blue), 11.5 (squares;
  magenta) and 11.9 (circles; black) at redshifts $2.2<z<2.5$.  The
  data are from Steidel et al. (2010) (large solid circles; red) and
  Rakic et al. (2012) (large filled squares; green).}
\label{fig:ew_R12}
\end{figure}

For comoving distances outside 0.7~Mpc, all the model predictions in
the left panel are in good agreement, with little difference with halo
mass. At smaller displacements, the \texttt{GADGET-3} simulation with
star formation (G30sfnw) shows a systematically larger spread in the
median optical depths compared with the \texttt{ENZO} simulation
(E60\_1024). The absorption increases approximately monotonically with
halo mass. Generally good agreement is found in comparison with the
data of \citet{2012ApJ...751...94R}, except for the innermost point,
where the observations show higher absorption than the simulations.

The right panel shows a plateau in the absorption for Hubble flow
distances within $\pm1$~Mpc of the halo centres, and a decline
beyond. As in the left panel, the median optical depths in the
\texttt{GADGET-3} simulations exceed those of the \texttt{ENZO}
simulation, with the $10^{11.5}\MSun$ haloes with winds producing the
greatest absorption. Outside of 3~Mpc, the median optical depths are
in good agreement between all the models, as well as with the
measurements of \citet{2012ApJ...751...94R}. The measured values
further in, however, greatly exceed all the model predictions.

Similar agreement with the data for the median optical as a function
of impact parameter was found by \citet{2013MNRAS.433.3103R}, with
their hydrodynamical simulations also falling well short of the
measured median optical depths as a function of Hubble separation
along the lines of sight. They point out that the discrepancy is
difficult to assess statistically because the data are highly
correlated along the line of sight. Agreement improves when averaging
over increasingly larger impact parameters.  Rather than pursuing
further comparisons with median optical depth measurements, however,
we turn next to integrated spectral quantities across wider
line-of-sight velocity intervals.  We find that these provide a more
direct means of discriminating between models.

\subsection{The velocity integrated Ly$\alpha$ equivalent width}

\citet{2012ApJ...751...94R} provide absorption equivalent widths
within $1000\kms$ wide velocity windows centred on the systemic
velocities of the galaxies in their sample. In Fig.~\ref{fig:ew_R12},
the equivalent widths (see Eq.~\ref{eq:ewDv}) within a spectral window
of width $\Delta v=1000\kms$ centred on the halo velocities from the
60~Mpc \texttt{ENZO} simulation (E60\_1024) and the \texttt{GADGET-3}
star formation simulations both without (G30sfnw) and with (G30sfw)
supernova feedback are shown as a function of impact parameter
$b_\perp$, averaged over all haloes in halo mass bins centred at
$\log_{10}M_h=11.1$, 11.5 and 11.9 of width
$\Delta\log_{10}M=0.2$. The simulated spectra have been adjusted by
applying a correction factor of 0.804 to the fluxes.  This matches the
correction applied by \citet{2012ApJ...751...94R} to allow for errors
in the continuum level in the earlier, lower resolution data of
\citet{2010ApJ...717..289S}. The equivalent width values rise towards
smaller impact parameters. At $b_\perp>0.4$~Mpc (comoving) there is
little difference between the model predictions. At all impact
parameters, the \texttt{ENZO} predictions are nearly independent of
halo mass, with the results for the $10^{11.1}\,M_\odot$ mass bin
slightly smaller than the others. By contrast, the \texttt{GADGET-3}
models show a systematic increase with halo mass for
$M_h>10^{11.1}\MSun$ within the inner $0.4$~Mpc. For reference, the
virial radius ranges over $150-330$~kpc for $10^{11}-10^{12}\,M_\odot$
haloes.\footnote{The comoving virial radius is
  $r_v\simeq0.33M_{h,12}^{1/3}$~Mpc for a halo mass
  $10^{12}M_{h,12}\,\MSun$. The turnaround radius, where the infall
  peculiar velocity cancels the Hubble flow, is at $r_{\rm
    t.a.}\simeq6r_v$ (Paper I).} The models including supernova
feedback predict somewhat larger equivalent widths compared with the
same mass halo with star formation alone. This produces a degeneracy
in the predictions:\ a given equivalent width value may be produced
either by a halo with star formation and no wind or a lower mass halo
with star formation and a wind.

Comparison with the equivalent width measurements of
\citet{2010ApJ...717..289S} (large red filled circles) and
\citet{2012ApJ...751...94R} (large green filled squares) shows
excellent agreement with the model predictions outside the virial
radius. The differences between the feedback and non-feedback models
are too small for the data to discriminate between
them. \citet{2012ApJ...752...39T} estimate a median halo mass of
$10^{11.9}\MSun$ for the survey, consistent with the findings here.

\subsection{Deviation of \Lya\ absorption from the mean IGM}

\begin{figure*}
\scalebox{0.5}{\includegraphics{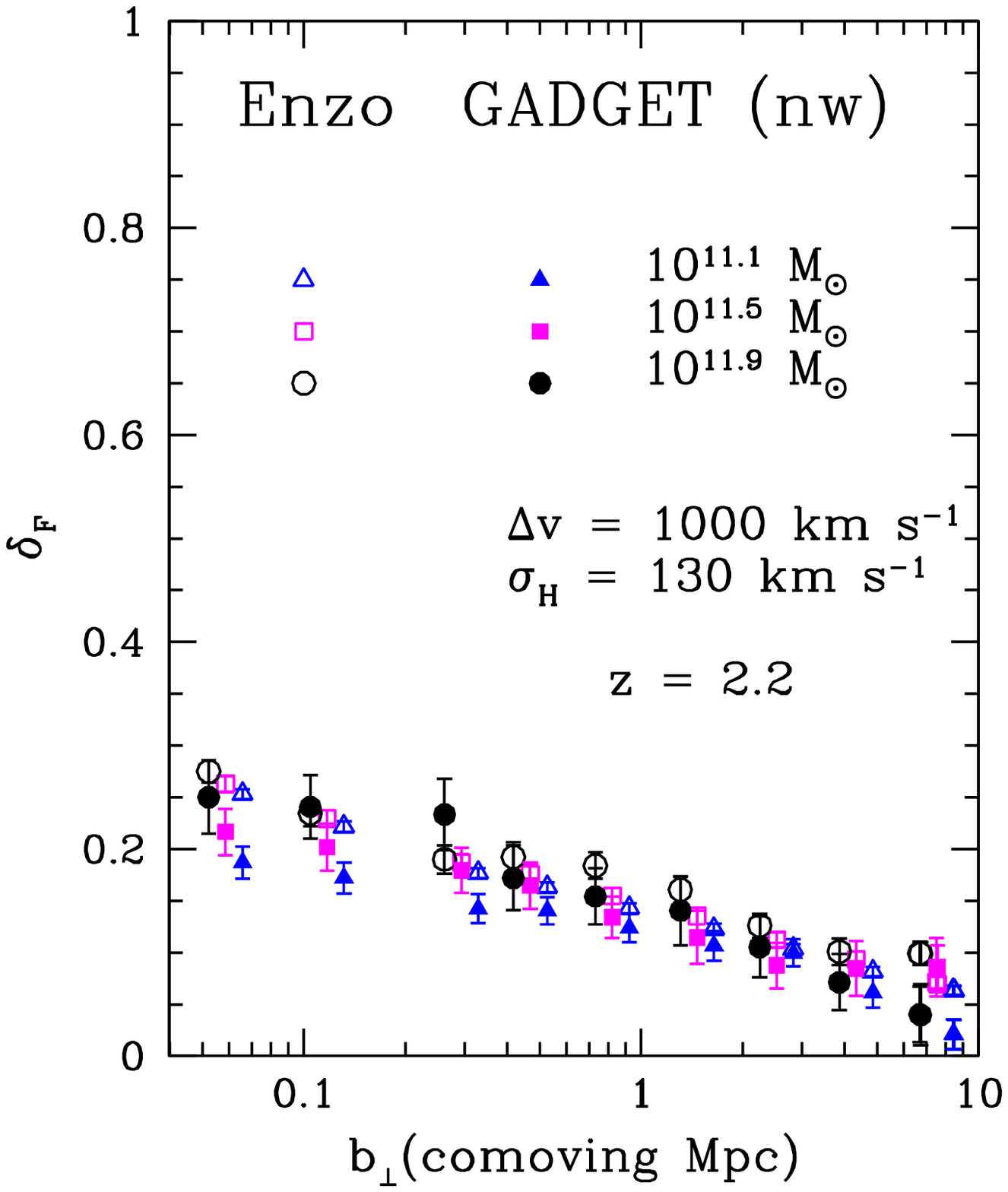}\includegraphics{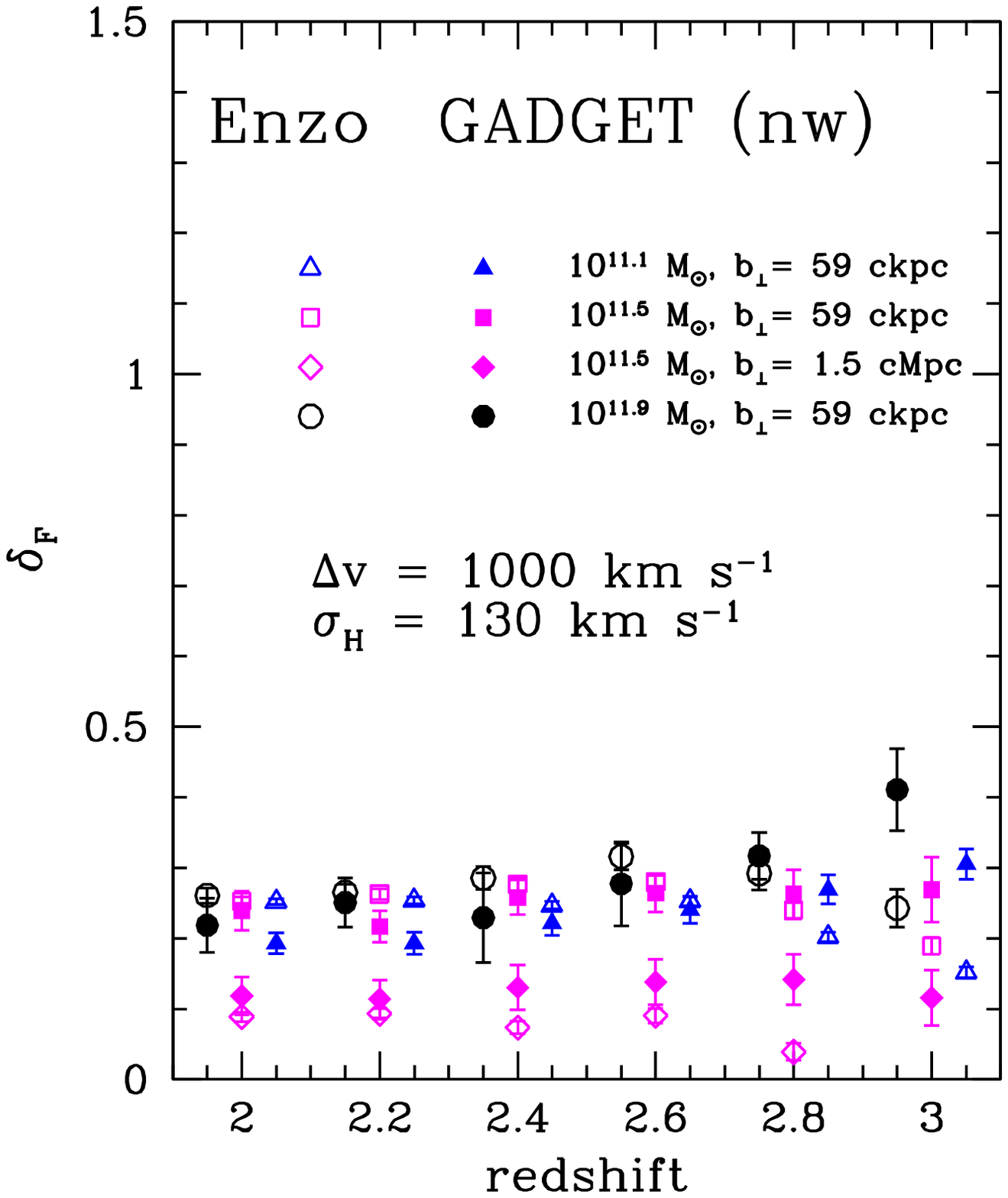}}
\vspace{-1.5cm}
\caption{Fractional absorption excess $\delta_F$ relative to the mean
  IGM absorption, for a spectral window $\Delta v=1000\kms$ wide
  across the halo systemic redshifts for the \texttt{ENZO} 60~Mpc box
  simulation (E60\_1024) and the \texttt{GADGET-3} simulation with
  star formation but no wind (G30sfnw). {\it Left panel:} The
  variation of $\delta_F$ with projected impact parameter $b_\perp$
  and halo mass at $z=2.2$.  Absorption exceeds the average IGM
  contribution out to several comoving Mpc. {\it Right:} The evolution
  of $\delta_F$ with redshift for selected values of $b_\perp$ and
  halo mass.  The symbols are as indicated on the diagram.  In both
  panels, halo velocities include a random component drawn from a
  Gaussian distribution with standard deviation $\sigma_{\rm
    H}=130\kms$.  }
\label{fig:df_comp_enzo60Mpc_gnw}
\end{figure*}

The absorption excess $\delta_F$ relative to the mean IGM, as
quantified by the effective optical depth $\tau_{\rm eff}$, averaged
over a spectral window $\Delta v=1000\kms$ at $z=2.2$, is shown in
Fig.~\ref{fig:df_comp_enzo60Mpc_gnw}.  Results are displayed for the
\texttt{ENZO} 60~Mpc box (E60\_1024) and the \texttt{GADGET-3}
simulation with star formation but no winds (G30sfnw), averaged over
all haloes in mass bins centred at $\log_{10}M_h=11.1$, 11.5 and 11.9
of width $\Delta\log_{10}M=0.2$.  At projected separations
$b_\perp>300$~kpc (comoving), all the models agree, showing a rise
towards the halo centres (left panel). The non-zero values rising
towards the halo centres demonstrate excess absorption above the
average IGM value in an extended region outside the haloes, well
outside the turnaround radii of the haloes ($r_{\rm t.a.}\sim1.8$~Mpc
comoving for $M_h=10^{11.9}\MSun$). While the results for the
\texttt{ENZO} simulation show essentially no variation with halo mass,
the absorption systematically rises within the virial radius with halo
mass in the \texttt{GADGET-3} simulations for haloes more massive than
$10^{11.1}\MSun$. At projected separations smaller than the virial
radius, the absorption in the \texttt{GADGET-3} simulation lies below
the \texttt{ENZO} values for halo masses below $10^{11.6}\,\MSun$. We
note that star formation in the \texttt{GADGET-3} simulation
substantially reduces the gas density within the virial radius
compared with the gas-conserving \texttt{ENZO} simulations (Paper I).

The excesses increase slowly with redshift, as shown in
the right panel of Fig.~\ref{fig:df_comp_enzo60Mpc_gnw}, rising by
only about 50 per cent from $z=2$ to 3. The trend of increasing
absorption with halo mass near the halo centres weakens for the
\texttt{GADGET-3} models with increasing redshift, except that the
absorption for the $10^{11.9}\MSun$ mass bin tends to stay the
largest. The different models also continue to show the same level of
absorption at large distances, here shown by the $10^{11.5}\MSun$
haloes at $b_\perp=1.5$Mpc.

\begin{figure*}
\scalebox{0.5}{\includegraphics{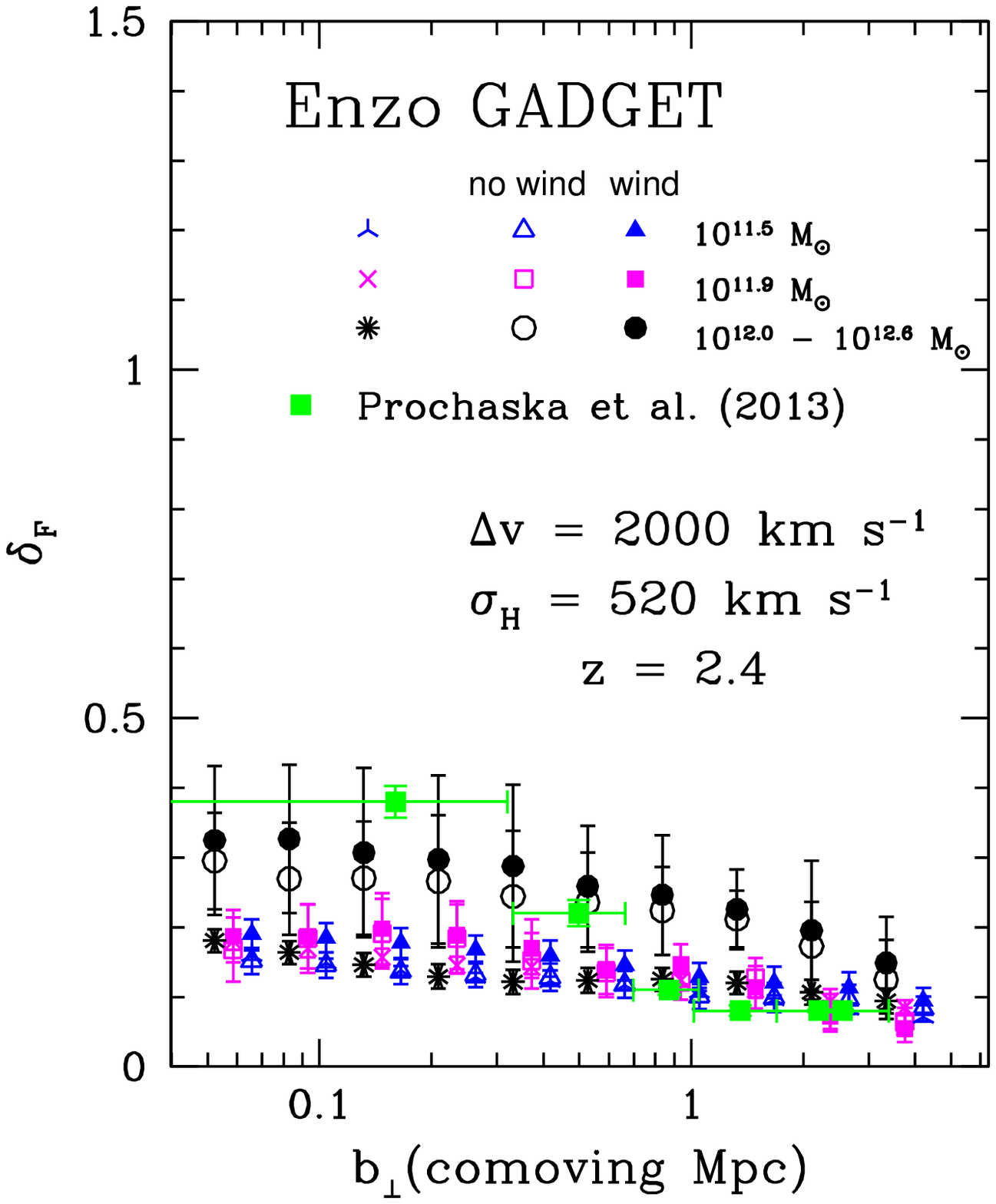}\includegraphics{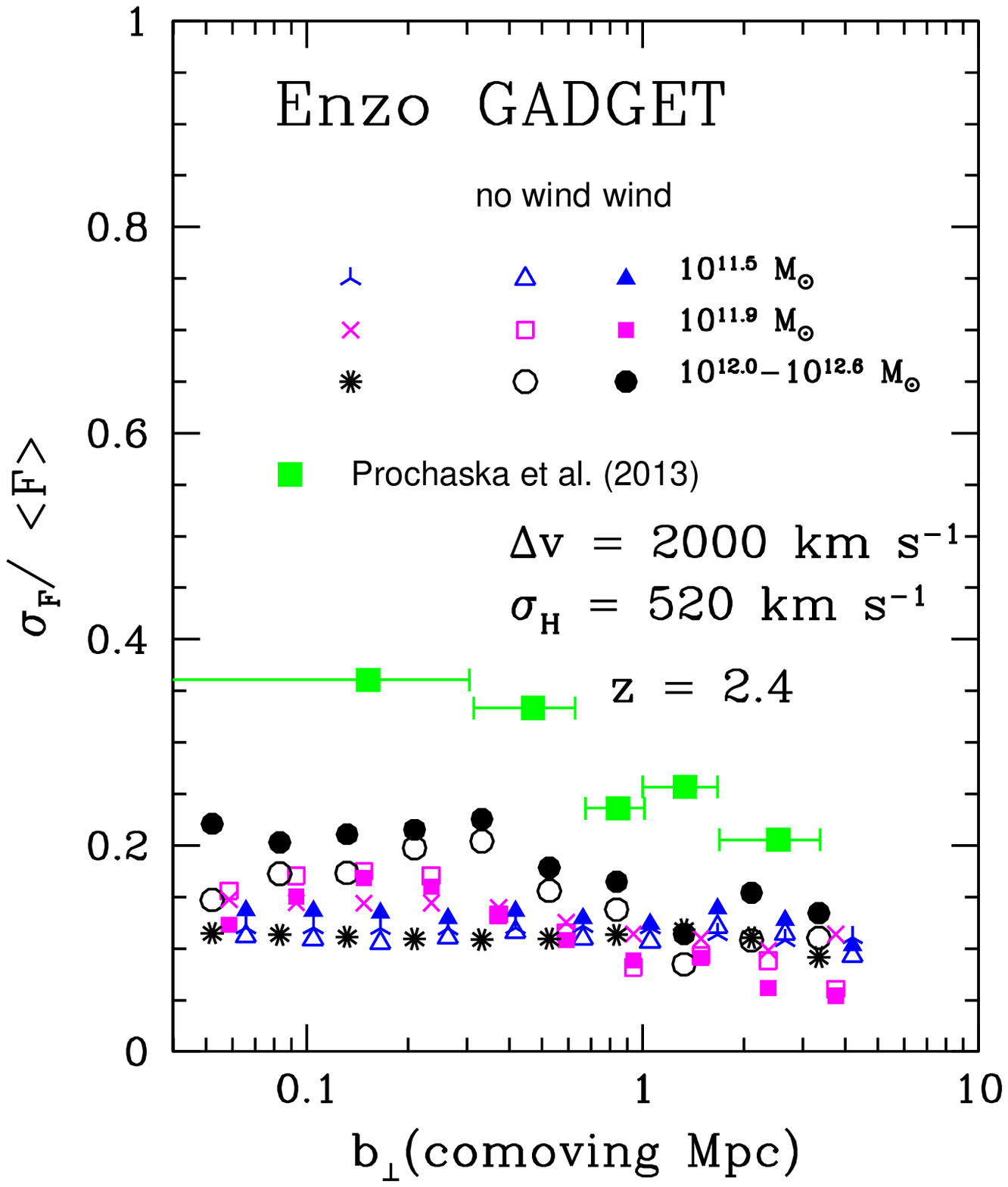}}
\vspace{-1.5cm}
\caption{{\it Left panel:} Fractional absorption excess $\delta_F$
  relative to the mean IGM absorption, for a spectral window $\Delta
  v=2000\kms$ across the halo systemic velocities for the
  \texttt{ENZO} 60~Mpc box simulation (E60\_1024) and the
  \texttt{GADGET-3} simulations with (G30sfw) and without (G30sfnw) a
  wind.  The data points in the left panel are for absorption around
  QSOs taken from \citet{2013ApJ...776..136P}, with the error bars
  showing the errors in the mean. {\it Right panel:} The fluctuations
  relative to the mean transmitted flux, $\sigma_F/\langle F\rangle$
  (see text for details).  Both $\delta_F$ and $\sigma_F/ \langle
  F\rangle$ are shown against projected impact parameter $b_\perp$ at
  $z=2.4$. The halo velocities include a random component drawn from a
  Gaussian distribution with standard deviation $\sigma_{\rm
    H}=520\kms$.}
\label{fig:df_sigf_comp_Enzo60_gnw_nw_P13}
\end{figure*}

Thus far we have compared the simulations with \Lya\ absorption data
around star-forming galaxies.  In the left panel of
Fig.~\ref{fig:df_sigf_comp_Enzo60_gnw_nw_P13} we instead compare with
the $\delta_F$ measurements surrounding QSOs reported by
\citet{2013ApJ...776..136P} at $z\approx2.4$, with a velocity window
$\Delta v=2000\kms$.  A random component drawn from a Gaussian
distribution with $\sigma_{\rm H}=520\kms$ is added to the halo
velocities to match the typical errors in the measured halo
redshifts. The predictions for the $M_h>10^{12}\,M_\odot$ haloes in
the \texttt{ENZO} simulation lie lower than those for the
\texttt{GADGET-3} simulations within the turnaround radius,
$r_{\rm t.a.}\sim 1.8$ Mpc, although the smaller box for the
\texttt{GADGET-3} simulations results in larger uncertainties. The
measurements outside the virial radius agree with all the models. The
innermost point, however, is consistent only with the
$M_h>10^{12}M_\odot$ halo predictions from the \texttt{GADGET-3}
simulations, although the uncertainties are large because of the small
number of haloes.

The fluctuations in the mean transmitted flux, given by
$\sigma_F=\langle(F-\langle F\rangle)^2\rangle^{1/2}$, are shown in
the right panel of Fig.~\ref{fig:df_sigf_comp_Enzo60_gnw_nw_P13} at
$z=2.4$. The ratio $\sigma_F/\langle F\rangle$ is nearly constant at
$0.10-0.15$ with $b_\perp$ for all the models. This is substantially
smaller than the measured fluctuations, especially within the
turnaround radius. Although the reported values include continuum
errors, so that they may perhaps be conservatively taken as upper
limits, the discrepancy may indicate the need for additional
astrophysical effects which increase the variance in transmitted flux
between lines of sight, such as QSO beaming.

\begin{figure*}
\scalebox{0.5}{\includegraphics{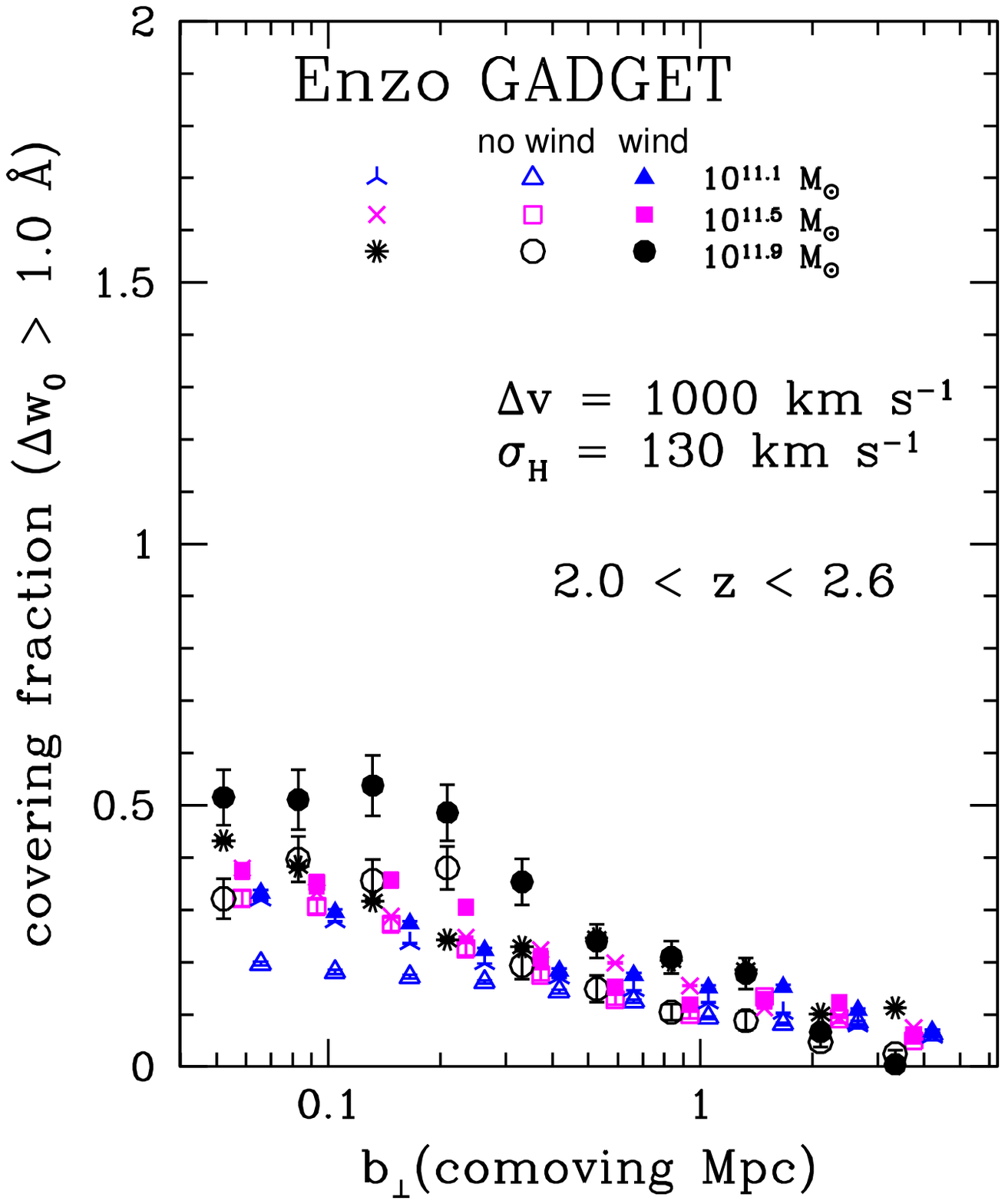}\includegraphics{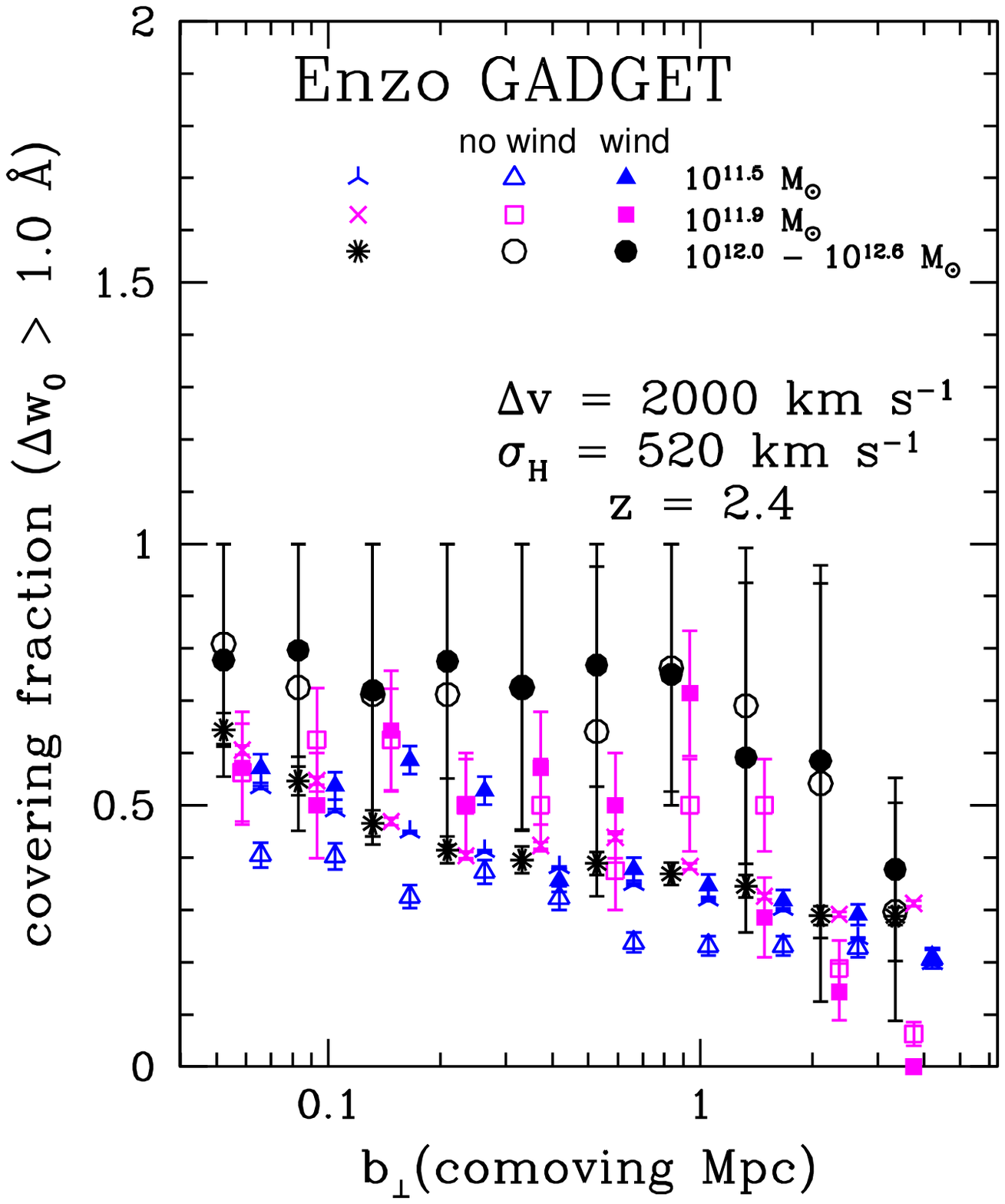}}
\vspace{-1.5cm}
\caption{Covering fraction $f_C$ for excess restframe equivalent
  widths of $\Delta w_0>1.0\rm\,\AA$ above the mean intergalactic
  value over spectral windows of width $\Delta v=1000\kms$ (left panel) and
  $\Delta v=2000\kms$ (right panel) across the halo systemic
  velocities for the \texttt{ENZO} 60~Mpc box (E60\_1024) and the
  \texttt{GADGET-3} simulations with (G30sfw) and without (G30sfnw) a
  wind. The covering fraction is shown against projected impact
  parameter $b_\perp$ at $2.0<z<2.6$ in the left panel, and at $z=2.4$
  in the right panel. The rise towards small impact parameters
  increases with increasing halo mass for the \texttt{GADGET-3}
  simulations. Within the virialized region, the models with a wind
  generally lie systematically above those without. The halo
  velocities include a random component drawn from a Gaussian
  distribution with standard deviation $\sigma_{\rm H}=130\kms$ (left
  panel) or $\sigma_{\rm H}=520\kms$ (right panel).}
\label{fig:fcov_comp_enzo_gadget_velw1000_2000}
\end{figure*}

Finally, a further independent test of the models is provided by
estimating the covering fractions $f_C$ for high equivalent width
absorption.  We define an excess integrated equivalent width relative
to the mean IGM as
\begin{equation}
  \Delta w_0(b_\perp, \Delta v) = \frac{\lambda_\alpha}{c}\int_{v_{\rm
      halo}-\Delta v/2}^{v_{\rm halo}+\Delta v/2}\,dv
  \left[e^{-\tau_{\rm eff}}-e^{-\tau_\alpha(b_\perp, v)}\right],
\label{eq:DewDv}
\end{equation}
where $\tau_{\rm eff}$ is the IGM effective optical depth. The
covering fraction is then the ratio of the number of lines of sight
with $\Delta w_0$ above a given threshold to the total number of lines
of sight.

\begin{figure*}
\scalebox{0.5}{\includegraphics{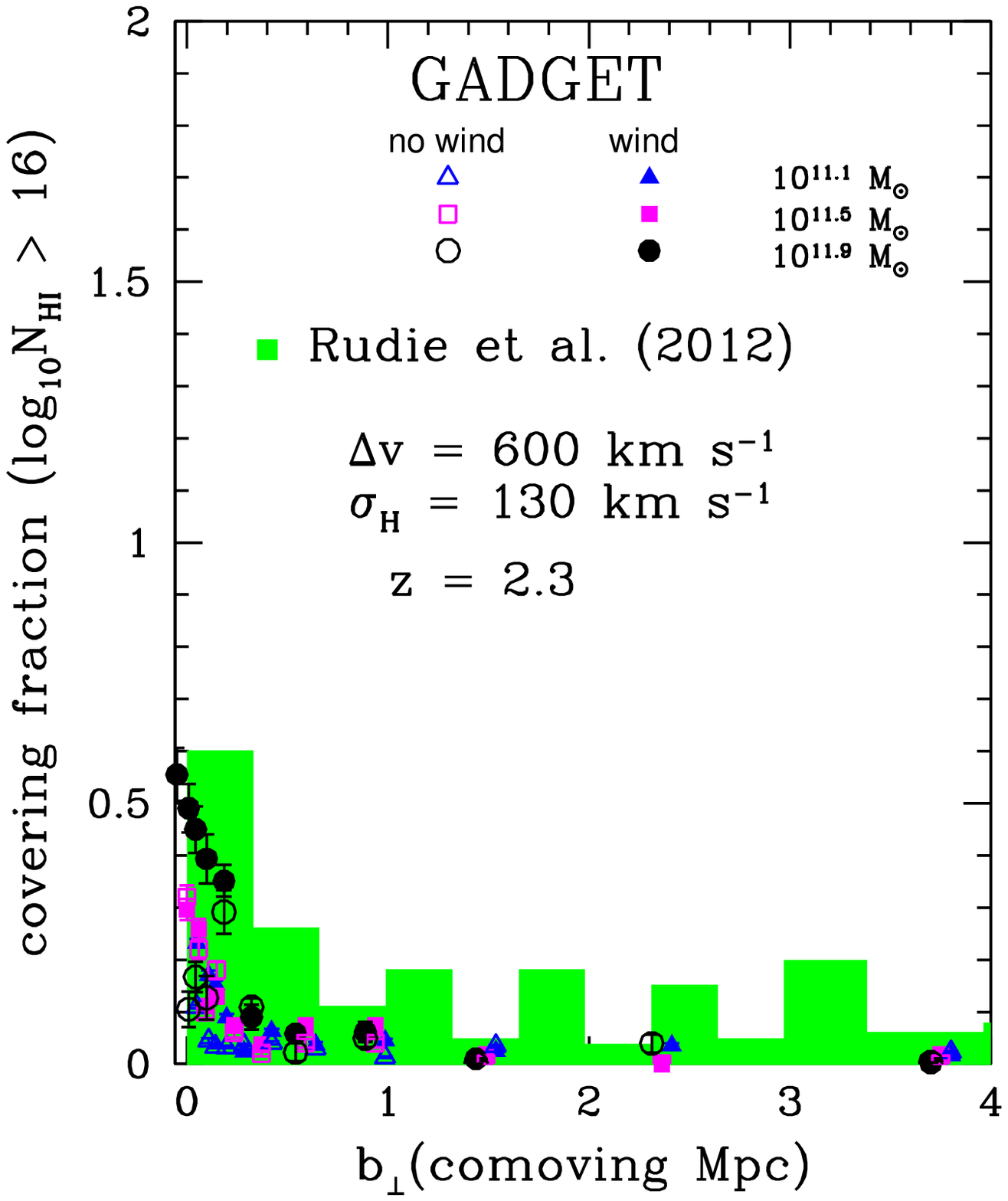}\includegraphics{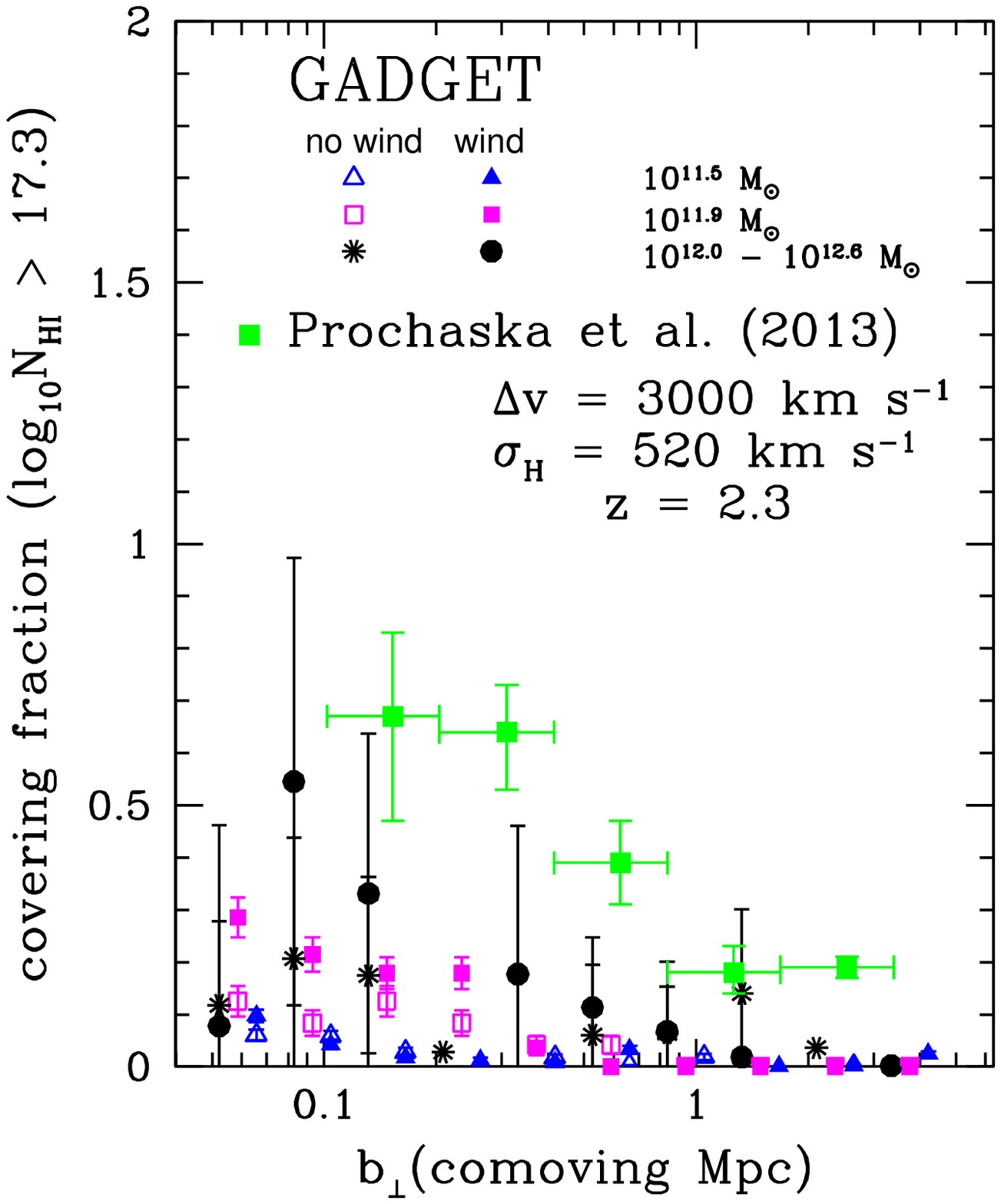}}
\vspace{-1.5cm}
\caption{Covering fraction $f_C$ for discrete absorption systems
  with column densities $N_{\rm HI}>10^{16}\rm\,cm^{-2}$ within a
  spectral window of width $\Delta v=600\kms$ across the halo systemic
  velocities (left panel) and $N_{\rm HI}>10^{17.3}\rm\,cm^{-2}$ for a
  window $\Delta v=3000\kms$ (right panel) for the \texttt{GADGET-3}
  simulations with (G30sfw) and without (G30sfnw) a wind.  The
  covering fraction is shown against projected impact parameter
  $b_\perp$ at $z=2.3$. The halo velocities include a random component
  drawn from a Gaussian distribution with standard deviation
  $\sigma_{\rm H}=130\kms$ (left panel) or $\sigma_{\rm H}=520\kms$
  (right panel).  }
\label{fig:fcov_abs_comp_gadget_nowind_wind}
\end{figure*}

The results for $\Delta w_0>1.0\rm\,\AA$ are shown in
Fig.~\ref{fig:fcov_comp_enzo_gadget_velw1000_2000} for the
\texttt{ENZO} 60~Mpc box (E60\_1024) and the \texttt{GADGET-3}
simulations with (G30sfw) and without (G30sfnw) winds. The left panel
is for absorption in the velocity window $\Delta v=1000\kms$ at
$2.0<z<2.6$, and the right for $\Delta v=2000\kms$ at $z=2.4$. The
covering fractions rise toward the halo centres in all the
models. Within the virial radius, the \texttt{GADGET-3} models show an
increase with halo mass and generally systematically higher values for
the models including wind feedback, especially for the narrower
velocity window in the left panel. The more massive haloes with winds
achieve covering fractions $f_C>0.5$ well within their virial radii.

While the integrated excess equivalent widths are straightforward to
measure from observations, the focus of recent literature has been the
covering fractions of identified discrete absorption systems. A direct
comparison with the simulations would require absorption line
identification and fitting, which we defer to later work. Here, we use
the high spatial resolution of the \texttt{GADGET-3} simulations to
provide illustrative comparisons by identifying discrete contiguous
\HI\ systems along the lines of sight.  In
Fig.~\ref{fig:fcov_abs_comp_gadget_nowind_wind}, we compare the
predicted covering fractions with the observations of
\citet{2012ApJ...750...67R} for systems with \HI\ column densities
$N_{\rm HI}>10^{16}\,{\rm cm^{-2}}$ within a velocity window $\Delta
v=600\kms$ wide centred on the galaxy systemic velocities (left
panel), and with the data of \citet{2013ApJ...776..136P} for
``optically thick'' absorbers in a velocity window $\Delta v=3000\kms$
wide centred on the galaxies (right panel). These authors define
optically thick absorbers variously as those showing obvious damping
wings, Lyman limit absorption, strong low-ionization metal absorption
or (if not classifiable otherwise) exhibiting a single strong
Ly$\alpha$ feature with $w_0>1.8\rm\,\AA$. We select discrete systems
from the simulations with column densities $N_{\rm HI}>10^{17.3}\,{\rm
  cm^{-2}}$.

The covering fraction of systems with $N_{\rm HI}>10^{16}\rm\,cm^{-2}$
is small compared with the measurements for most of the models,
particularly within the virial radius. The exception is for massive
haloes, $M_h>10^{11.8}\,\MSun$, for which the model with a wind shows
a rapid rise in the covering fraction toward the halo centre close to
the measured values. The covering fractions for absorbers optically
thick at the Lyman limit ($N_{\rm HI}>10^{17.3}\rm\,cm^{-2}$) lie
systematically below the observations of
\citet{2013ApJ...776..136P}. While the covering fraction for
$M_h>10^{12}\,\MSun$ approaches the measured values within the spread
from the simulation, the uncertainties in the mean are large due to
the small number of haloes available in this mass bin (a total of
12). This, along with the heterogeneous definition of the observed
optically thick absorbers, makes it difficult to quantify
statistically the level of disagreement. Moreover,
\citet{2013ApJ...776..136P} caution against errors due to continuum
placement and line-blending. Similarly low covering fractions compared
with observations are found by \citet{2014ApJ...780...74F} and
\citet{2014arXiv1409.1919F} for halo masses exceeding
$10^{12}\,\MSun$. By contrast, a recent study by
\citet{2015arXiv150305553R} finds close agreement with the measured
values, which they attribute to their feedback model. Their
comparison, however, is for larger mass haloes,
$M_h>10^{12.5}\,\MSun$, captured within their larger simulation
volume. It should also be noted that the simulations other than ours
compute the covering fractions based on integrated line-of-sight
column densities rather than discrete absorption systems.

\section{Discussion}
\label{sec:discussion}

\subsection{Radial characteristics of the absorption}
\label{subsec:absorption}

We have compared the statistics for \Lya\ absorption along lines of
sight passing through the environments of galaxy haloes using three
simulations:\ an \texttt{ENZO} simulation without star formation and
two \texttt{GADGET-3} simulations with star formation, one of which
also allows for feedback in the form of supernovae-driven winds. We
found in Paper I that the gas properties of the haloes largely agree
beyond the turnaround radius, regardless of the inclusion of star
formation or winds. The results presented here confirm that the
predicted absorption line statistics for lines of sight passing
outside the turnaround radius are similar.

Comparisons with the \Lya\ absorption statistics in SFGs
\citep{2010ApJ...717..289S, 2012ApJ...751...94R} and QSOs
\citep{2013ApJ...776..136P}, demonstrate the
simulations successfully recover the absorption statistics outside the
halo virialized regions. Within the virial radius, however, the
measured amount of integrated absorption rises to a level none of the models
reproduce. A similar trend is found using the median Ly$\alpha$
optical depth data for SFGs (Fig.~\ref{fig:taumed_comp_enzo_gadget}).
  
The various absorption line statistics nevertheless show some common
trends among the models. The predictions of all the models, whether or
not they include star formation or wind feedback, all agree, within
the error bars, outside the turnaround radii of the haloes. The models
also agree well with the data for both SFGs and QSOs in this
region. As the virial radius is approached, however, the models begin
to diverge. 

The gas around galaxies has generally been divided between a
circumgalactic medium within a distance of $\sim300$~kpc (proper) from
a galaxy \citep[e.g.][]{2010ApJ...717..289S, 2013ApJ...776..136P}, and
the IGM on larger scales. Based on our simulation results, we rather
suggest it is useful to consider three distinct regions:\ the inner
virialized region, an intermediate region we call the \lq mesogalactic
medium' (MGM), extending between the virial radius and the IGM
($r_v<r<12r_v$), and the IGM well outside galactic haloes
($r>12r_v$). The virialized region is distinguished as the active
theatre within which star formation and feedback most affect the
hydrogen absorption signatures. Here, discrete optically thick
absorption systems, especially damped \Lya\ absorbers, may make
substantial contributions to the integrated amount of absorption. The
MGM is the region around galaxies exhibiting enhanced absorption over
the diffuse IGM, but little affected by star formation or
feedback. It originates from gas in the extended haloes around
galaxies (Paper I), before merging into the diffuse IGM on larger
scales.

\subsection{From where does the measured absorption arise?}
\label{subsec:location}

Many of the \HI\ absorption observational studies around galaxies have
focussed on the search for evidence of the cold streams predicted to
penetrate into the virialized regions of moderate mass galactic haloes
($M_h<10^{12}\MSun$). Establishing that the absorption arises from
inward streaming gas is observationally challenging; indeed, the
measured kinematics favour outflows \citep{2010ApJ...717..289S}. The
measured covering fractions of cold ($T\approx10^4$~K) gas appear to
exceed theoretical expectations in both SFGs
\citep{2012ApJ...751...94R} and QSOs \citep{2013ApJ...776..136P}. The
excess absorption in QSOs is especially intriguing, as the gas
interior to the virial radius is expected to be shock heated to such
high temperatures as to be in collisional ionization equilibrium, with
greatly reduced levels of neutral hydrogen. As shown in
Fig.~\ref{fig:halo_gadget_z2_los}, the absorption signal as a function
of impact parameter is complicated by the complex peculiar velocity
field of the gas. The large peculiar motions of the gas in the
vicinity of galaxies can shift the absorption contribution of gas
parcels distant from the galaxies into and out of the velocity window
used to measure the absorption signature.

\begin{figure*}
\scalebox{0.5}{\includegraphics{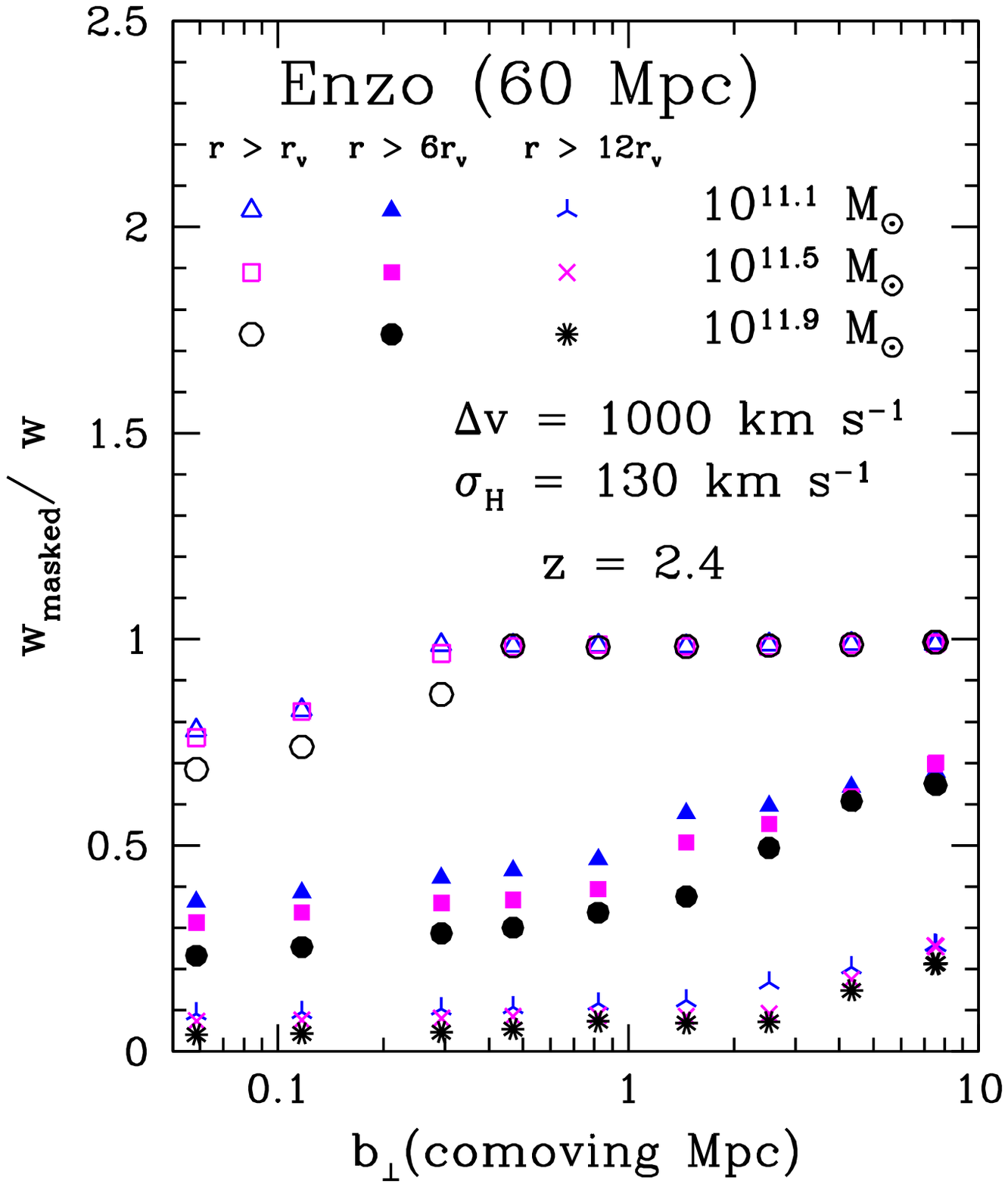}\includegraphics{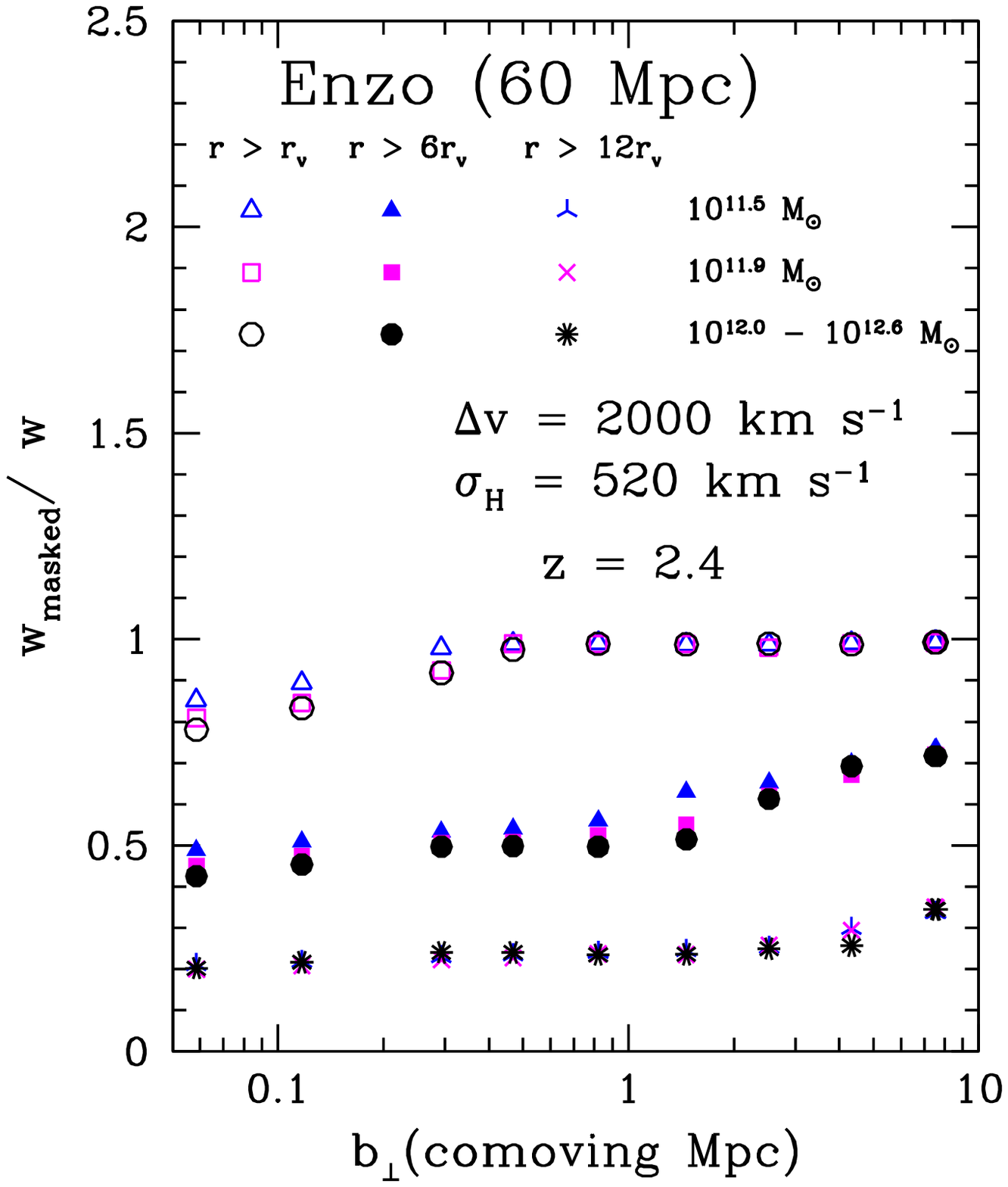}}
\vspace{-1.5cm}
\caption{Fraction of equivalent width arising from gas outside the
  virialized region, $r>r_v$ (open symbols), outside the turnaround
  radius, $r>6r_v$ (solid symbols), and outside the mesogalactic
  region, $r>12r_v$ (starred symbols), shown against projected
  comoving impact parameter $b_\perp$ at $z=2.4$. The equivalent
  widths are computed within spectral windows of width
  $\Delta v=1000\kms$ (left panel) or $2000\kms$ (right panel) across
  the halo systemic velocities for the \texttt{ENZO} 60~Mpc box
  (E60\_1024). In both panels, most of the absorption for lines of
  sight passing within the virial radius arises from gas outside the
  virialized region, and about a third from gas outside the turnaround
  radius. The halo velocities include a random component drawn from a
  Gaussian distribution with standard deviation
  $\sigma_{\rm H}=130\kms$ (left panel) or $520\kms$ (right panel).  }
\label{fig:ew_comp_Enzo_nowind_maskrv_bcom}
\end{figure*} 

In order to gain some insight into the origin of the measured
absorption signatures, we construct spectra from the \texttt{ENZO} 60
Mpc simulation after masking all the \HI\ within the virial radius of
haloes more massive than $M_h>10^{11}\MSun$. The resulting effect on
the equivalent width in a velocity window $\Delta v=1000\kms$ wide
centred on the galaxy velocities is shown in the left panel of
Fig.~\ref{fig:ew_comp_Enzo_nowind_maskrv_bcom} (open symbols). While
the contribution of gas external to the virialized region to the
absorption signature decreases towards small impact parameters for
$b_\perp<r_v$, it never vanishes. At least 70 per cent of the
absorption at $b_\perp << r_v$ arises from gas outside the virialized
zone.

In terms of the absolute equivalent width values,
Fig.~\ref{fig:ew_R12} shows that within the virial radius the
simulations underpredict the measured equivalent width by $\Delta
w_0\simeq1\rm\,\AA$ compared with observations. About half the mean
absorption is thus missing within the virial radius. Zoom-in
simulations, achieving a baryon mass resolution of about
$10^5\,M_\odot$ for haloes more massive than $10^{11}\,M_\odot$, find
covering fractions for saturated absorption lines smaller than 50 per
cent \citep[e.g.][]{2011MNRAS.418.1796F, 2014ApJ...780...74F,
  2012MNRAS.424.2292G}. Until damping wings appear ($N_{\rm
  HI}>10^{19}\,{\rm cm}^{-2}$), an individual saturated line will
contribute only about $\Delta w_0\simeq0.5\rm\,\AA$. Given the
covering fraction, this is too small to make up the
difference. Systems with column densities $N_{\rm HI}>10^{19}\,{\rm
  cm}^{-2}$, or equivalent widths of $2-3\rm\,\AA$, are rare, with
covering fractions well under 10 per cent in these simulations, so
that including systems showing damping wings cannot fully make up the
difference either. The simulations of \citet{2014arXiv1409.1919F},
with a baryon resolution as small as $4\times10^4\,M_\odot$, however,
show a higher incidence rate of damped absorbers. For haloes more
massive than $10^{11}\,M_\odot$, the covering fractions for saturated
lines reach 60--80 per cent, and 20--40 per cent for systems with
$N_{\rm HI}>10^{19}\,{\rm cm}^{-2}$. This may just be sufficient to
make up the missing absorption from the virialized zone. If the
relatively rare systems with damped wings account for the missing
absorption, then wide variations in the integrated absorption between
different lines of sight is expected, with the absorption found in the
simulations presented here providing a more constant baseline level.

Masking out the region within the turnaround radius ($r<6r_v$) further
reduces the signal, but the contribution to absorption within $b_\perp
< 6r_v$ from material outside the turnaround radius is still
non-negligible. Gas at $r>6r_v$ contributes at least 25 per cent of the
absorption signal we obtain even at $b_\perp<<r_v$ (by comparison,
less than 10 per cent of the signal derives from the diffuse IGM, at
$r>12r_v$). At transverse separations outside the turnaround radius,
the full absorption signal is still not achieved, demonstrating that
about 20 per cent of the absorption along a line of sight passing
through the mesogalactic region of a given halo arises from
mesogalactic gas in other haloes. For a halo of mass
$M_h=10^{11.9}\MSun$, this corresponds to absorption from material at
a distance beyond 1.8~Mpc (comoving), or a velocity offset of
$130\kms$, well within the velocity window $\Delta v=1000\kms$.

Increasing the velocity window to $\Delta v=2000\kms$
(Fig.~\ref{fig:ew_comp_Enzo_nowind_maskrv_bcom}, right panel) produces
similar trends. Results for halo masses corresponding to the host
galaxies of QSOs ($12.0<\log_{10} M_h<12.6$) are now also shown. For the
increased velocity window, even for these massive haloes 80 per cent
of the absorption signal at small transverse impact parameters
($b_\perp << r_v$), originates in gas outside the virialized
region. About half the signal for lines of sight passing within the
turnaround radius originates from gas beyond the turnaround radius,
while 20 per cent of the absorption for lines of sight passing through
the virialized and mesogalactic regions originates from gas in the
diffuse IGM.

\section{Conclusions}
\label{sec:conclusions}

We perform large-scale cosmological hydrodynamical simulations using
two numerical schemes, \texttt{GADGET-3}, including star formation
using the prescription of \citet{2003MNRAS.339..289S}, with and
without supernovae-driven wind feedback, and \texttt{ENZO} without
star formation as a control case. The simulations allow us to assess
the impact of star formation and wind feedback separately on the
\HI\ absorption statistics around galaxies and QSOs. The statistical
measures we consider are based primarily on the integrated absorption
properties within velocity windows centred on the systemic velocities
of the galactic haloes:\ equivalent width values, fractional
absorption excesses relative to the mean IGM and fluctuations in the
absorbed flux. We also provide statistical predictions for covering
fractions of integrated equivalent width values over the velocity
windows and for some discrete absorption systems. We compare the
simulation results with the measurements of
\citet{2010ApJ...717..289S}, \citet{2012ApJ...751...94R} and
\citet{2012ApJ...750...67R} for star-forming galaxies and
\citet{2013ApJ...776..136P} for QSOs over the redshift range $2<z<3$.

On the basis of the simulated absorption properties, we identify three
regions in the gas surrounding a galaxy with distinct absorption
properties, the inner virialized region, a mesogalactic zone extending
from the virial radius to twice the turnaround radius ($r_{\rm
  t.a.}\simeq 6r_v\approx1-3$~cMpc for $10^{11.0}-10^{12.6}\MSun$
haloes), and the IGM beyond. The simulations show star formation and
wind feedback play only a secondary role in establishing the
integrated \HI\ absorption signatures compared with the general trend
of increasing absorption for decreasing impact parameter. All the
simulations reproduce the absorption measurements within the
mesogalactic medium and beyond. This is a non-trivial result since it
extends the success of IGM simulations in the context of a
$\Lambda$CDM cosmology, from the diffuse IGM to the extended haloes of
galaxies down to their virial radii, without being subject to the
uncertainties of subgrid physics. As such, the agreement is a
confirmation of the $\Lambda$CDM cosmological model for structure
formation on comoving scales down to $\sim0.4$~Mpc around galaxies
over the redshift range $2<z<3$.

The \texttt{GADGET-3} simulations, both with and without a wind, and
the \texttt{ENZO} 60~Mpc box simulation produce comparable equivalent
width values in $\Delta v=1000\kms$ windows centred on the halo
systemic velocities over the full range of impact parameters. Allowing
for a wind, all the model predictions for haloes with masses
$M_h>10^{11.4}\MSun$ agree with the measurements of
\citet{2010ApJ...717..289S} and \citet{2012ApJ...751...94R} outside
the virial radius. Within the virialized region, the \texttt{GADGET-3}
models show a secondary trend of increasing absorption with halo mass
and enhanced absorption for a given halo mass when wind feedback is
included, producing a degeneracy between halo mass and wind feedback,
at least for the feedback model we adopt. All the models, however,
underpredict the amount of absorption compared with observations for
lines of sight passing through the virialized regions. High resolution
zoom-in simulations including radiative transfer suggest Lyman limit
systems, especially damped \Lya\ absorbers, may account for the
remaining absorption.

We also compare our models with the median optical depth measurements
of \citet{2012ApJ...751...94R}. The models agree well with the
measurements outside the virial radius, where all the models show
similar values, but not within, similar to the integrated equivalent
width comparisons. In addition to possibly being a consequence of
unresolved absorption systems, the shortfall may also be in part
statistical, a consequence of line-of-sight correlations in the median
optical depth values on small scales \citep[we refer the reader
to][for a discussion]{2013MNRAS.433.3103R}.

We also compare the model predictions for the fractional absorption
excess $\delta_F$ with the data of \citet{2013ApJ...776..136P}, who
report measured values in velocity windows $\Delta v=2000\kms$ centred
on QSOs. In the mesogalactic region, both the \texttt{ENZO} and
\texttt{GADGET-3} simulation predictions are largely insensitive to
the halo mass. Agreement with the measured values is achieved even for
halo masses below $M_h>10^{12.0}\MSun$, the expected lower halo mass
of QSO host galaxies. The \texttt{ENZO} simulation underpredicts the
amount of absorption within the virial radius, as found for SFGs. The
comparison with the \texttt{GADGET-3} simulations is inconclusive
because of the low number of massive haloes in the simulation volume.

We compute covering fractions for integrated equivalent widths within
the velocity windows, as these are readily measured from observations
without the requirement of identifying and fitting discrete absorption
systems in the spectra. The sensitivity we find to feedback suggests
the covering fractions may provide useful constraints on feedback
models.  Since much of the literature focusses instead on the covering
fraction of discrete absorption systems, we also consider illustrative
comparisons with observations for saturated absorption lines. We find
good agreement with observations of SFGs, but underpredict the
covering fraction of optically thick absorbers in QSOs, with only
marginal agreement with the observations. Because of the low number of
massive haloes in our simulations, however, the statistics are too
poor to be conclusive. Agreement may require larger mass haloes than
may be examined with the limited box size analysed here.  An accurate
determination of the contributions from individual Lyman limit systems
and damped Ly$\alpha$ absorbers to the absorption properties within
the virialized region may also require the increased resolution of
zoom-in simulations.  Recovering the full amount of absorption from
virialized gas will likely require additional physical effects,
including alternative sub-grid feedback models, self-consistent
radiative hydrodynamics to account for the physical response of
systems optically thick to ionizing radiation, or possibly pressure
resulting from a fluctuating magnetic field
\citep{2013ApJ...762...15P, 2014MNRAS.437.3639C}.

\section*{Acknowledgments}
This work used the DiRAC Data Analytic system at the University of
Cambridge, operated by the University of Cambridge High Performance
Computing Service on behalf of the STFC DiRAC HPC Facility
(www.dirac.ac.uk). This equipment was funded by BIS National
E-infrastructure capital grant (ST/K001590/1), STFC capital grants
ST/H008861/1 and ST/H00887X/1, and STFC DiRAC Operations grant
ST/K00333X/1. DiRAC is part of the National
E-Infrastructure. Additional computations were performed on facilities
funded by an STFC Rolling-Grant and consolidated grant. JSB
acknowledges the support of a Royal Society University Research
Fellowship. ERT is supported by an STFC consolidated grant. We thank
V. Springel for making \texttt{GADGET-3} available. Computations
described in this work were performed using the {\texttt{ENZO}} code
developed by the Laboratory for Computational Astrophysics at the
University of California in San Diego (http://lca.ucsd.edu).

\bibliographystyle{mn2e-eprint}
\bibliography{apj-jour,lya}

\begin{thebibliography}{}

\bibitem[\protect\citeauthoryear{{Abazajian}, {Adelman-McCarthy},
  {Ag{\"u}eros}, {Allam}, {Allende Prieto}, {An}, {Anderson}, {Anderson},
  {Annis}, {Bahcall} \& et al.}{{Abazajian} et~al.}{2009}]{2009ApJS..182..543A}
{Abazajian} K.~N.,  {Adelman-McCarthy} J.~K.,  {Ag{\"u}eros} M.~A.,  {Allam}
  S.~S.,  {Allende Prieto} C.,  {An} D.,  {Anderson} K.~S.~J.,  {Anderson}
  S.~F.,  {Annis} J.,  {Bahcall} N.~A.,    et al. 2009, \apjs, 182, 543

\bibitem[\protect\citeauthoryear{{Ahn}, {Alexandroff}, {Allende Prieto},
  {Anderson}, {Anderton}, {Andrews}, {Aubourg}, {Bailey}, {Balbinot}, {Barnes}
  \& et al.}{{Ahn} et~al.}{2012}]{2012ApJS..203...21A}
{Ahn} C.~P.,  {Alexandroff} R.,  {Allende Prieto} C.,  {Anderson} S.~F.,
  {Anderton} T.,  {Andrews} B.~H.,  {Aubourg} {\'E}.,  {Bailey} S.,  {Balbinot}
  E.,  {Barnes} R.,    et al. 2012, \apjs, 203, 21

\bibitem[\protect\citeauthoryear{{Becker}, {Bolton}, {Haehnelt} \&
  {Sargent}}{{Becker} et~al.}{2011}]{2011MNRAS.410.1096B}
{Becker} G.~D.,  {Bolton} J.~S.,  {Haehnelt} M.~G.,    {Sargent} W.~L.~W.,
  2011, \mnras, 410, 1096

\bibitem[\protect\citeauthoryear{{Becker}, {Hewett}, {Worseck} \&
  {Prochaska}}{{Becker} et~al.}{2013}]{2013MNRAS.430.2067B}
{Becker} G.~D.,  {Hewett} P.~C.,  {Worseck} G.,    {Prochaska} J.~X.,  2013,
  \mnras, 430, 2067

\bibitem[\protect\citeauthoryear{{Birnboim} \& {Dekel}}{{Birnboim} \&
  {Dekel}}{2003}]{2003MNRAS.345..349B}
{Birnboim} Y.,  {Dekel} A.,  2003, \mnras, 345, 349

\bibitem[\protect\citeauthoryear{{Bryan}, {Norman}, {O'Shea}, {Abel}, {Wise},
  {Turk}, {Reynolds}, {Collins}, {Wang}, {Skillman}, {Smith} \& {$+$}}{{Bryan}
  et~al.}{2014}]{2014ApJS..211...19B}
{Bryan} G.~L.,  {Norman} M.~L.,  {O'Shea} B.~W.,  {Abel} T.,  {Wise} J.~H.,
  {Turk} M.~J.,  {Reynolds} D.~R.,  {Collins} D.~C.,  {Wang} P.,  {Skillman}
  S.~W.,  {Smith} B.,    {$+$} 2014, \apjs, 211, 19

\bibitem[\protect\citeauthoryear{{Chongchitnan} \& {Meiksin}}{{Chongchitnan} \&
  {Meiksin}}{2014}]{2014MNRAS.437.3639C}
{Chongchitnan} S.,  {Meiksin} A.,  2014, \mnras, 437, 3639

\bibitem[\protect\citeauthoryear{{Crighton}, {Bielby}, {Shanks}, {Infante},
  {Bornancini}, {Bouch{\'e}}, {Lambas}, {Lowenthal}, {Minniti}, {Morris},
  {Padilla}, {P{\'e}roux}, {Petitjean}, {Theuns} \& {$+$}}{{Crighton}
  et~al.}{2011}]{2011MNRAS.414...28C}
{Crighton} N.~H.~M.,  {Bielby} R.,  {Shanks} T.,  {Infante} L.,  {Bornancini}
  C.~G.,  {Bouch{\'e}} N.,  {Lambas} D.~G.,  {Lowenthal} J.~D.,  {Minniti} D.,
  {Morris} S.~L.,  {Padilla} N.,  {P{\'e}roux} C.,  {Petitjean} P.,  {Theuns}
  T.,    {$+$} 2011, \mnras, 414, 28

\bibitem[\protect\citeauthoryear{{Croft}}{{Croft}}{2004}]{2004ApJ...610..642C}
{Croft} R.~A.~C.,  2004, \apj, 610, 642

\bibitem[\protect\citeauthoryear{{Dalla Vecchia} \& {Schaye}}{{Dalla Vecchia}
  \& {Schaye}}{2008}]{2008MNRAS.387.1431D}
{Dalla Vecchia} C.,  {Schaye} J.,  2008, \mnras, 387, 1431

\bibitem[\protect\citeauthoryear{{Dekel} \& {Birnboim}}{{Dekel} \&
  {Birnboim}}{2006}]{2006MNRAS.368....2D}
{Dekel} A.,  {Birnboim} Y.,  2006, \mnras, 368, 2

\bibitem[\protect\citeauthoryear{{Dekel} \& {Silk}}{{Dekel} \&
  {Silk}}{1986}]{1986ApJ...303...39D}
{Dekel} A.,  {Silk} J.,  1986, \apj, 303, 39

\bibitem[\protect\citeauthoryear{{Faucher-Giguere}, {Hopkins}, {Keres},
  {Muratov}, {Quataert} \& {Murray}}{{Faucher-Giguere}
  et~al.}{2014}]{2014arXiv1409.1919F}
{Faucher-Giguere} C.-A.,  {Hopkins} P.~F.,  {Keres} D.,  {Muratov} A.~L.,
  {Quataert} E.,    {Murray} N.,  2014, ArXiv e-prints, 1409.1919

\bibitem[\protect\citeauthoryear{{Faucher-Gigu{\`e}re} \& {Kere{\v
  s}}}{{Faucher-Gigu{\`e}re} \& {Kere{\v s}}}{2011}]{2011MNRAS.412L.118F}
{Faucher-Gigu{\`e}re} C.-A.,  {Kere{\v s}} D.,  2011, \mnras, 412, L118

\bibitem[\protect\citeauthoryear{{Fumagalli}, {Hennawi}, {Prochaska}, {Kasen},
  {Dekel}, {Ceverino} \& {Primack}}{{Fumagalli}
  et~al.}{2014}]{2014ApJ...780...74F}
{Fumagalli} M.,  {Hennawi} J.~F.,  {Prochaska} J.~X.,  {Kasen} D.,  {Dekel} A.,
   {Ceverino} D.,    {Primack} J.,  2014, \apj, 780, 74

\bibitem[\protect\citeauthoryear{{Fumagalli}, {Prochaska}, {Kasen}, {Dekel},
  {Ceverino} \& {Primack}}{{Fumagalli} et~al.}{2011}]{2011MNRAS.418.1796F}
{Fumagalli} M.,  {Prochaska} J.~X.,  {Kasen} D.,  {Dekel} A.,  {Ceverino} D.,
   {Primack} J.~R.,  2011, \mnras, 418, 1796

\bibitem[\protect\citeauthoryear{{Genzel}, {Newman}, {Jones}, {F{\"o}rster
  Schreiber}, {Shapiro}, {Genel}, {Lilly}, {Renzini}, {Tacconi} \&
  {$+$}}{{Genzel} et~al.}{2011}]{2011ApJ...733..101G}
{Genzel} R.,  {Newman} S.,  {Jones} T.,  {F{\"o}rster Schreiber} N.~M.,
  {Shapiro} K.,  {Genel} S.,  {Lilly} S.~J.,  {Renzini} A.,  {Tacconi} L.~J.,
   {$+$} 2011, \apj, 733, 101

\bibitem[\protect\citeauthoryear{{Goerdt}, {Dekel}, {Sternberg}, {Gnat} \&
  {Ceverino}}{{Goerdt} et~al.}{2012}]{2012MNRAS.424.2292G}
{Goerdt} T.,  {Dekel} A.,  {Sternberg} A.,  {Gnat} O.,    {Ceverino} D.,  2012,
  \mnras, 424, 2292

\bibitem[\protect\citeauthoryear{{Haardt} \& {Madau}}{{Haardt} \&
  {Madau}}{2012}]{2012ApJ...746..125H}
{Haardt} F.,  {Madau} P.,  2012, \apj, 746, 125

\bibitem[\protect\citeauthoryear{{Haas}, {Schaye}, {Booth}, {Dalla Vecchia},
  {Springel}, {Theuns} \& {Wiersma}}{{Haas} et~al.}{2013}]{2013MNRAS.435.2931H}
{Haas} M.~R.,  {Schaye} J.,  {Booth} C.~M.,  {Dalla Vecchia} C.,  {Springel}
  V.,  {Theuns} T.,    {Wiersma} R.~P.~C.,  2013, \mnras, 435, 2931

\bibitem[\protect\citeauthoryear{{Harrison}, {Alexander}, {Mullaney} \&
  {Swinbank}}{{Harrison} et~al.}{2014}]{2014MNRAS.441.3306H}
{Harrison} C.~M.,  {Alexander} D.~M.,  {Mullaney} J.~R.,    {Swinbank} A.~M.,
  2014, \mnras, 441, 3306

\bibitem[\protect\citeauthoryear{{Hernquist}, {Katz}, {Weinberg} \&
  {Miralda-Escud{\'e}}}{{Hernquist} et~al.}{1996}]{1996ApJ...457L..51H}
{Hernquist} L.,  {Katz} N.,  {Weinberg} D.~H.,    {Miralda-Escud{\'e}} J.,
  1996, \apjl, 457, L51

\bibitem[\protect\citeauthoryear{{Hinshaw}, {Larson}, {Komatsu}, {Spergel},
  {Bennett}, {Dunkley}, {Nolta}, {Halpern} \& {et al.}}{{Hinshaw}
  et~al.}{2013}]{2013ApJS..208...19H}
{Hinshaw} G.,  {Larson} D.,  {Komatsu} E.,  {Spergel} D.~N.,  {Bennett} C.~L.,
  {Dunkley} J.,  {Nolta} M.~R.,  {Halpern} M.,    {et al.} 2013, \apjs, 208, 19

\bibitem[\protect\citeauthoryear{{Kay}, {Pearce}, {Frenk} \& {Jenkins}}{{Kay}
  et~al.}{2002}]{2002MNRAS.330..113K}
{Kay} S.~T.,  {Pearce} F.~R.,  {Frenk} C.~S.,    {Jenkins} A.,  2002, \mnras,
  330, 113

\bibitem[\protect\citeauthoryear{{Kay}, {Pearce}, {Jenkins}, {Frenk}, {White},
  {Thomas} \& {Couchman}}{{Kay} et~al.}{2000}]{2000MNRAS.316..374K}
{Kay} S.~T.,  {Pearce} F.~R.,  {Jenkins} A.,  {Frenk} C.~S.,  {White} S.~D.~M.,
   {Thomas} P.~A.,    {Couchman} H.~M.~P.,  2000, \mnras, 316, 374

\bibitem[\protect\citeauthoryear{{Kere{\v s}}, {Katz}, {Weinberg} \&
  {Dav{\'e}}}{{Kere{\v s}} et~al.}{2005}]{2005MNRAS.363....2K}
{Kere{\v s}} D.,  {Katz} N.,  {Weinberg} D.~H.,    {Dav{\'e}} R.,  2005,
  \mnras, 363, 2

\bibitem[\protect\citeauthoryear{{Kirkman} \& {Tytler}}{{Kirkman} \&
  {Tytler}}{2008}]{2008MNRAS.391.1457K}
{Kirkman} D.,  {Tytler} D.,  2008, \mnras, 391, 1457

\bibitem[\protect\citeauthoryear{{Larson}}{{Larson}}{1974}]{1974MNRAS.169..229L}
{Larson} R.~B.,  1974, \mnras, 169, 229

\bibitem[\protect\citeauthoryear{{Meiksin}, {Bolton} \& {Tittley}}{{Meiksin}
  et~al.}{2014}]{2014MNRAS.445.2462M}
{Meiksin} A.,  {Bolton} J.~S.,    {Tittley} E.~R.,  2014, \mnras, 445, 2462 (Paper I)

\bibitem[\protect\citeauthoryear{{Meiksin}}{{Meiksin}}{2009}]{2009RvMP...81.1405M}
{Meiksin} A.~A.,  2009, Reviews of Modern Physics, 81, 1405

\bibitem[\protect\citeauthoryear{{Miralda-Escud{\'e}}}{{Miralda-Escud{\'e}}}{2005}]{MiraldaEscude2005}
{Miralda-Escud{\'e}} J.,  2005, \apjl, 620, L91, astro-ph/0410315

\bibitem[\protect\citeauthoryear{{Nelson}, {Genel}, {Pillepich},
  {Vogelsberger}, {Springel} \& {Hernquist}}{{Nelson}
  et~al.}{2015}]{2015arXiv150302665N}
{Nelson} D.,  {Genel} S.,  {Pillepich} A.,  {Vogelsberger} M.,  {Springel} V.,
    {Hernquist} L.,  2015, ArXiv e-prints, 1503.02665

\bibitem[\protect\citeauthoryear{{Pandey} \& {Sethi}}{{Pandey} \&
  {Sethi}}{2013}]{2013ApJ...762...15P}
{Pandey} K.~L.,  {Sethi} S.~K.,  2013, \apj, 762, 15

\bibitem[\protect\citeauthoryear{{Prochaska}, {Hennawi}, {Lee}, {Cantalupo},
  {Bovy}, {Djorgovski}, {Ellison}, {Lau}, {Martin}, {Myers}, {Rubin} \&
  {Simcoe}}{{Prochaska} et~al.}{2013}]{2013ApJ...776..136P}
{Prochaska} J.~X.,  {Hennawi} J.~F.,  {Lee} K.-G.,  {Cantalupo} S.,  {Bovy} J.,
   {Djorgovski} S.~G.,  {Ellison} S.~L.,  {Lau} M.~W.,  {Martin} C.~L.,
  {Myers} A.,  {Rubin} K.~H.~R.,    {Simcoe} R.~A.,  2013, \apj, 776, 136

\bibitem[\protect\citeauthoryear{{Prochaska}, {Hennawi} \&
  {Simcoe}}{{Prochaska} et~al.}{2013}]{2013ApJ...762L..19P}
{Prochaska} J.~X.,  {Hennawi} J.~F.,    {Simcoe} R.~A.,  2013, \apjl, 762, L19

\bibitem[\protect\citeauthoryear{{Rahmati}, {Pawlik}, {Raicevi\'c} \&
  {Schaye}}{{Rahmati} et~al.}{2013}]{2013MNRAS.430.2427R}
{Rahmati} A.,  {Pawlik} A.~H.,  {Raicevi\'c} M.,    {Schaye} J.,  2013, \mnras,
  430, 2427

\bibitem[\protect\citeauthoryear{{Rahmati}, {Schaye}, {Bower}, {Crain},
  {Furlong}, {Schaller} \& {Theuns}}{{Rahmati}
  et~al.}{2015}]{2015arXiv150305553R}
{Rahmati} A.,  {Schaye} J.,  {Bower} R.~G.,  {Crain} R.~A.,  {Furlong} M.,
  {Schaller} M.,    {Theuns} T.,  2015, ArXiv e-prints, 1503.05553

\bibitem[\protect\citeauthoryear{{Rakic}, {Schaye}, {Steidel}, {Booth}, {Dalla
  Vecchia} \& {Rudie}}{{Rakic} et~al.}{2013}]{2013MNRAS.433.3103R}
{Rakic} O.,  {Schaye} J.,  {Steidel} C.~C.,  {Booth} C.~M.,  {Dalla Vecchia}
  C.,    {Rudie} G.~C.,  2013, \mnras, 433, 3103

\bibitem[\protect\citeauthoryear{{Rakic}, {Schaye}, {Steidel} \&
  {Rudie}}{{Rakic} et~al.}{2012}]{2012ApJ...751...94R}
{Rakic} O.,  {Schaye} J.,  {Steidel} C.~C.,    {Rudie} G.~C.,  2012, \apj, 751,
  94

\bibitem[\protect\citeauthoryear{{Rudie}, {Steidel}, {Trainor}, {Rakic},
  {Bogosavljevi{\'c}}, {Pettini}, {Reddy}, {Shapley}, {Erb} \& {Law}}{{Rudie}
  et~al.}{2012}]{2012ApJ...750...67R}
{Rudie} G.~C.,  {Steidel} C.~C.,  {Trainor} R.~F.,  {Rakic} O.,
  {Bogosavljevi{\'c}} M.,  {Pettini} M.,  {Reddy} N.,  {Shapley} A.~E.,  {Erb}
  D.~K.,    {Law} D.~R.,  2012, \apj, 750, 67

\bibitem[\protect\citeauthoryear{{Schaye}}{{Schaye}}{2006}]{Schaye2006}
{Schaye} J.,  2006, \apj, 643, 59, astro-ph/0409137

\bibitem[\protect\citeauthoryear{{Schaye}, {Crain}, {Bower}, {Furlong},
  {Schaller}, {Theuns}, {Dalla Vecchia}, {Frenk} \& {et al.}}{{Schaye}
  et~al.}{2015}]{Schaye2015}
{Schaye} J.,  {Crain} R.~A.,  {Bower} R.~G.,  {Furlong} M.,  {Schaller} M.,
  {Theuns} T.,  {Dalla Vecchia} C.,  {Frenk} C.~S.,    {et al.} 2015, \mnras,
  446, 521

\bibitem[\protect\citeauthoryear{{Schaye}, {Dalla Vecchia}, {Booth}, {Wiersma},
  {Theuns}, {Haas}, {Bertone}, {Duffy}, {McCarthy} \& {van de Voort}}{{Schaye}
  et~al.}{2010}]{2010MNRAS.402.1536S}
{Schaye} J.,  {Dalla Vecchia} C.,  {Booth} C.~M.,  {Wiersma} R.~P.~C.,
  {Theuns} T.,  {Haas} M.~R.,  {Bertone} S.,  {Duffy} A.~R.,  {McCarthy} I.~G.,
     {van de Voort} F.,  2010, \mnras, 402, 1536

\bibitem[\protect\citeauthoryear{{Scholz} \& {Walters}}{{Scholz} \&
  {Walters}}{1991}]{1991ApJ...380..302S}
{Scholz} T.~T.,  {Walters} H.~R.~J.,  1991, \apj, 380, 302

\bibitem[\protect\citeauthoryear{{Shen}, {Madau}, {Guedes}, {Mayer},
  {Prochaska} \& {Wadsley}}{{Shen} et~al.}{2013}]{2013ApJ...765...89S}
{Shen} S.,  {Madau} P.,  {Guedes} J.,  {Mayer} L.,  {Prochaska} J.~X.,
  {Wadsley} J.,  2013, \apj, 765, 89

\bibitem[\protect\citeauthoryear{{Silk}, {Djorgovski}, {Wyse} \& {Bruzual
  A.}}{{Silk} et~al.}{1986}]{1986ApJ...307..415S}
{Silk} J.,  {Djorgovski} S.,  {Wyse} R.~F.~G.,    {Bruzual A.} G.,  1986, \apj,
  307, 415

\bibitem[\protect\citeauthoryear{{Springel}}{{Springel}}{2005}]{2005MNRAS.364.1105S}
{Springel} V.,  2005, \mnras, 364, 1105

\bibitem[\protect\citeauthoryear{{Springel} \& {Hernquist}}{{Springel} \&
  {Hernquist}}{2003}]{2003MNRAS.339..289S}
{Springel} V.,  {Hernquist} L.,  2003, \mnras, 339, 289

\bibitem[\protect\citeauthoryear{{Steidel}, {Erb}, {Shapley}, {Pettini},
  {Reddy}, {Bogosavljevi{\'c}}, {Rudie} \& {Rakic}}{{Steidel}
  et~al.}{2010}]{2010ApJ...717..289S}
{Steidel} C.~C.,  {Erb} D.~K.,  {Shapley} A.~E.,  {Pettini} M.,  {Reddy} N.,
  {Bogosavljevi{\'c}} M.,  {Rudie} G.~C.,    {Rakic} O.,  2010, \apj, 717, 289

\bibitem[\protect\citeauthoryear{{Suresh}, {Bird}, {Vogelsberger}, {Genel},
  {Torrey}, {Sijacki}, {Springel} \& {Hernquist}}{{Suresh}
  et~al.}{2015}]{2015MNRAS.448..895S}
{Suresh} J.,  {Bird} S.,  {Vogelsberger} M.,  {Genel} S.,  {Torrey} P.,
  {Sijacki} D.,  {Springel} V.,    {Hernquist} L.,  2015, \mnras, 448, 895

\bibitem[\protect\citeauthoryear{{Trainor} \& {Steidel}}{{Trainor} \&
  {Steidel}}{2012}]{2012ApJ...752...39T}
{Trainor} R.~F.,  {Steidel} C.~C.,  2012, \apj, 752, 39

\bibitem[\protect\citeauthoryear{{van de Voort}, {Schaye}, {Altay} \&
  {Theuns}}{{van de Voort} et~al.}{2012}]{2012MNRAS.421.2809V}
{van de Voort} F.,  {Schaye} J.,  {Altay} G.,    {Theuns} T.,  2012, \mnras,
  421, 2809

\bibitem[\protect\citeauthoryear{{van de Voort}, {Schaye}, {Booth} \& {Dalla
  Vecchia}}{{van de Voort} et~al.}{2011}]{2011MNRAS.415.2782V}
{van de Voort} F.,  {Schaye} J.,  {Booth} C.~M.,    {Dalla Vecchia} C.,  2011,
  \mnras, 415, 2782

\bibitem[\protect\citeauthoryear{{Viel}, {Haehnelt} \& {Springel}}{{Viel}
  et~al.}{2004}]{2004MNRAS.354..684V}
{Viel} M.,  {Haehnelt} M.~G.,    {Springel} V.,  2004, \mnras, 354, 684

\end{thebibliography}

\appendix

\section{Convergence tests on \HI\ statistics}

\begin{figure}
\scalebox{0.5}{\includegraphics{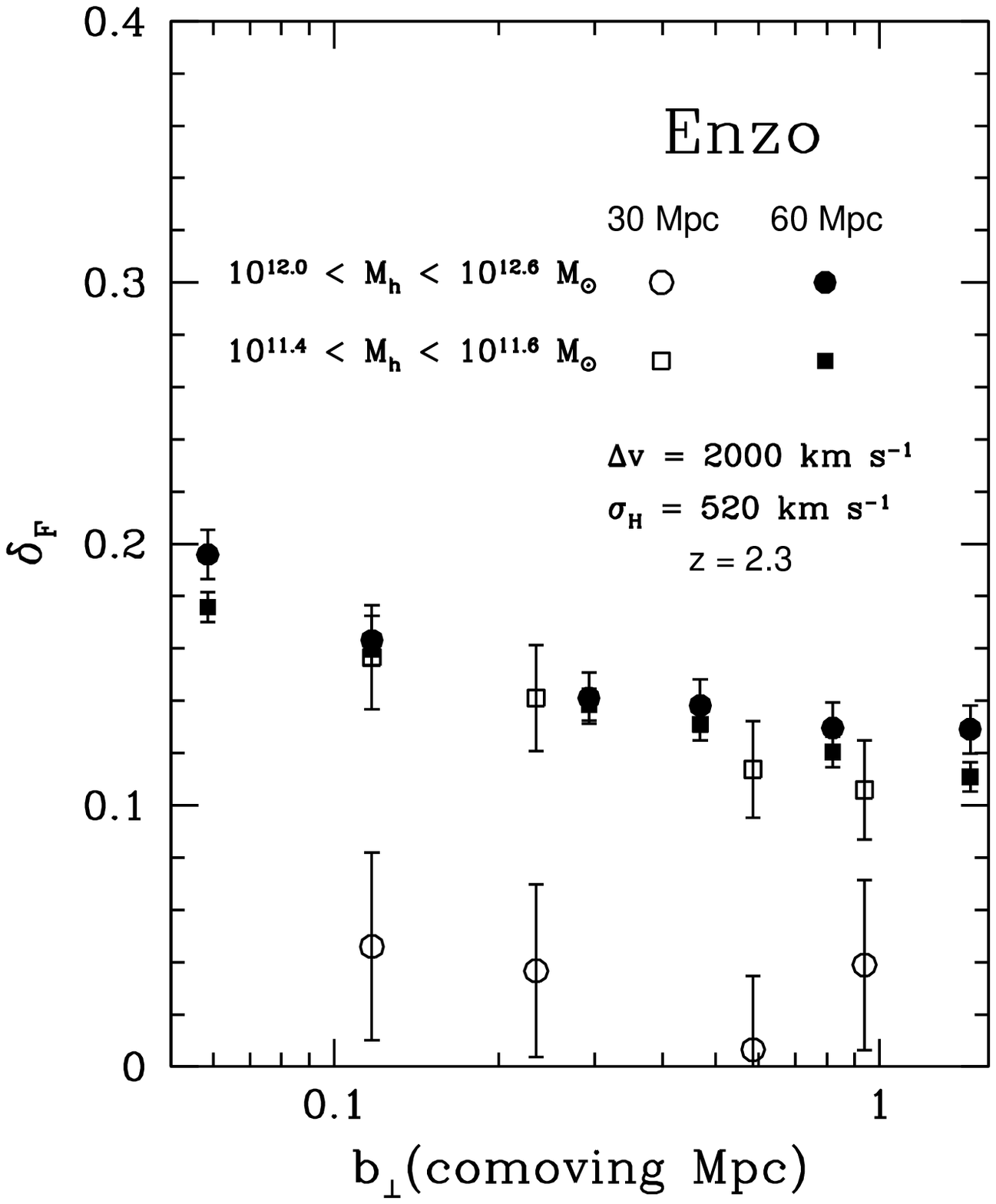}}
\vspace{-1.5cm}
\caption{Fractional absorption excess $\delta_F$ relative to the mean
  IGM absorption for a velocity window $\Delta v=2000\kms$ around the
  halo centre-of-mass velocity.  The data are shown as a function of
  the line-of-sight impact parameter $b_\perp$ for halo masses
  $11.4<\log_{10}M_h=11.6$ (squares) and $12.0<\log_{10}M_h<12.6$
  (circles) at $z=2.3$, for the \texttt{ENZO} simulations in boxes of
  side 30~Mpc (E30\_1024, open symbols) and 60~Mpc (E60\_1024, filled
  symbols). While the predictions converge well for moderate mass
  haloes, differences arise for the most massive haloes, although the
  uncertainties are large because of the small number of haloes in the
  30~Mpc box. Halo velocities include a random component drawn from a
  Gaussian distribution with standard deviation $\sigma_{\rm
    H}=520\kms$.  }
\label{fig:df_conv_enzo}
\end{figure}

The convergence of the fractional absorption excess $\delta_F$ with
box size is tested in Fig.~\ref{fig:df_conv_enzo} using the
\texttt{ENZO} simulations for a velocity window $\Delta v=2000\kms$
and a halo redshift uncertainty $\sigma_{\rm H}=520\kms$, similar to
the observations of \citet{2013ApJ...762L..19P}. Only small changes in
$\delta_F$, less than 0.02, are found for the gas around
$\sim10^{11.5}\,\MSun$ haloes. In the high mass bin there is
considerable Poisson scatter in $\delta_F$ from the smaller box, in
which there are only 14 haloes with $M_h>10^{12}\,M_\odot$.

\begin{figure}
\scalebox{0.5}{\includegraphics{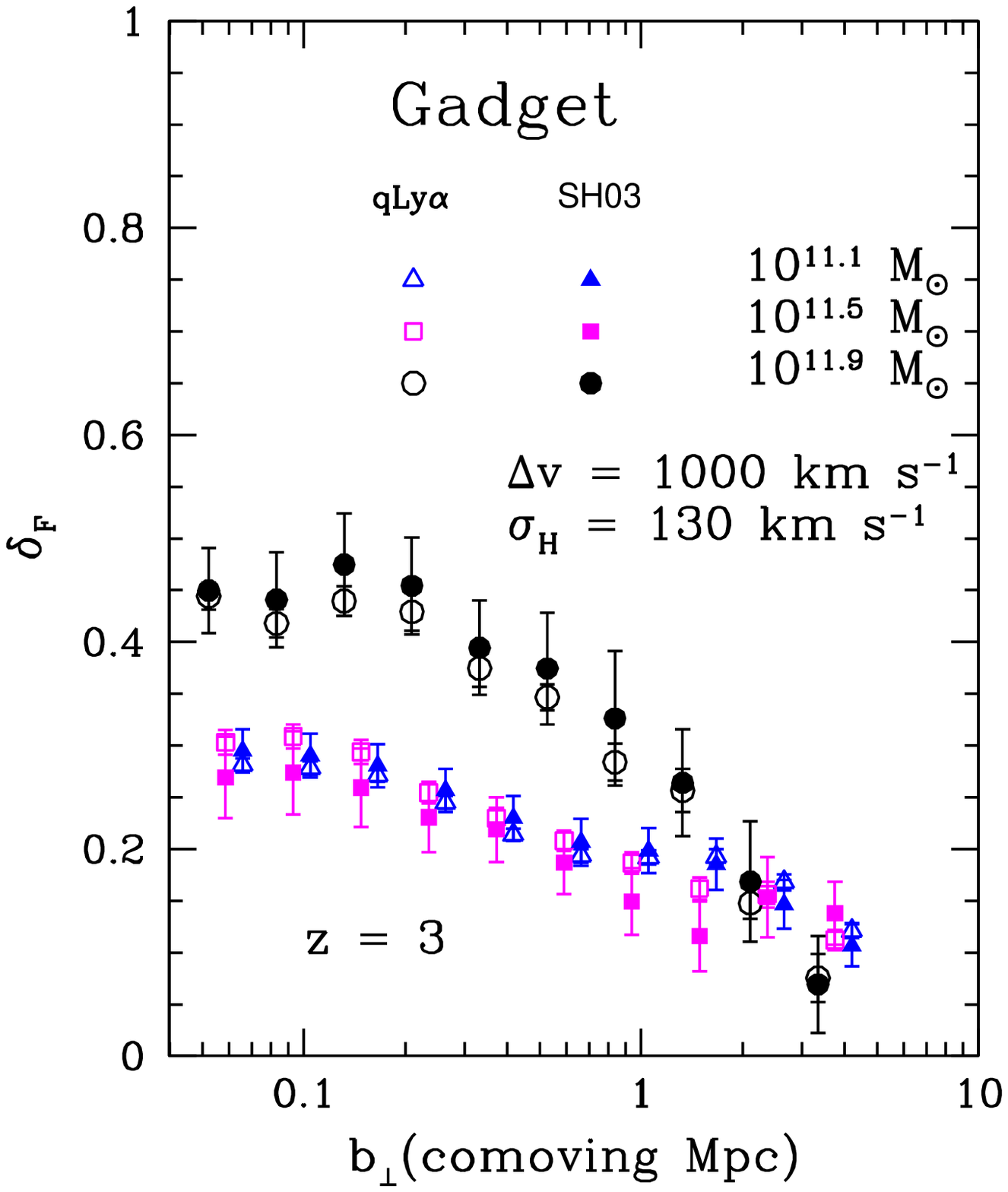}}
\vspace{-1.5cm}
\caption{Fractional absorption excess $\delta_F$ relative to the
  mean IGM absorption within a $\Delta v=1000\kms$ window centred on
  the halo centre-of-mass velocity.  The data are shown as a function
  of the line-of-sight impact parameter $b_\perp$ for halo masses
  $\log_{10}M_h=11.1$ (triangles; blue), 11.5 (squares; magenta) and
  11.9 (circles; black) at redshift $z=3$, for the \texttt{GADGET-3}
  non-wind simulations using the default star-formation prescription
  of SH03 (G39sfnw, filled symbols) or quick Ly$\alpha$
  (G30qLy$\alpha$ open symbols). The choice of gas removal algorithm
  has little impact on the amount of absorption. Halo velocities
  include a random component drawn from a Gaussian distribution with
  standard deviation $\sigma_{\rm H}=130\kms$.  }
\label{fig:df_comp_Gadget_nowind_qLya}
\end{figure}

We examine the possible role radiative transfer may have on the
absorption features using the simplifying approximation of an
attenuated radiation field within systems sufficiently dense to be
self-shielding to photoionizing radiation. We adopt the prescription
of \citet{2013MNRAS.430.2427R}, using a characteristic self-shielding
total hydrogen density of $0.0064T_4^{0.17}\,{\rm cm^{-3}}$ for
temperature $T_4=T/10^4\,{\rm K}$. The effect on the mean values of
$\delta_F$ are under 5 per cent, and generally less than one per
cent. We therefore neglect the effects of radiative transfer in this
paper on the integrated amount of absorption. Radiative transfer does
affect the column densities of discrete absorption systems, but acts
principally by creating systems with \HI\ column densities
sufficiently large to produce damping wings
($N_{\rm HI}>10^{19}\,{\rm cm^{-2}}$).  Since we cannot include the
physical response of the gas due to the resulting temperature and
consequent internal pressure changes these systems would undergo, we
do not consider them except for their small contribution to the lower
thresholds of systems we do examine,
$N_{\rm HI}>10^{16}\,{\rm cm^{-2}}$ and
$N_{\rm HI}>10^{17.3}\,{\rm cm^{-2}}$ in
Fig.~\ref{fig:fcov_abs_comp_gadget_nowind_wind}, for which the effects
of radiative transfer are included. We have also compared results for
$\delta_F$ using Doppler and Voigt profile functions for run
G30sfnw. The differences were negligible. We use the Doppler profile
function for the results presented here since it requires an order of
magnitude less analysis time. The Voigt function was used, however,
when including the effects of radiative transfer.

All the \texttt{GADGET-3} simulations for the results in this paper
use the star formation prescription of \citet{2003MNRAS.339..289S}. In
Fig.~\ref{fig:df_comp_Gadget_nowind_qLya} we compare the results for
$\delta_F$ using the quick Ly$\alpha$ method instead. Very little
difference is found, suggesting the results are robust to the method
of gas removal.

\begin{figure}
\scalebox{0.5}{\includegraphics{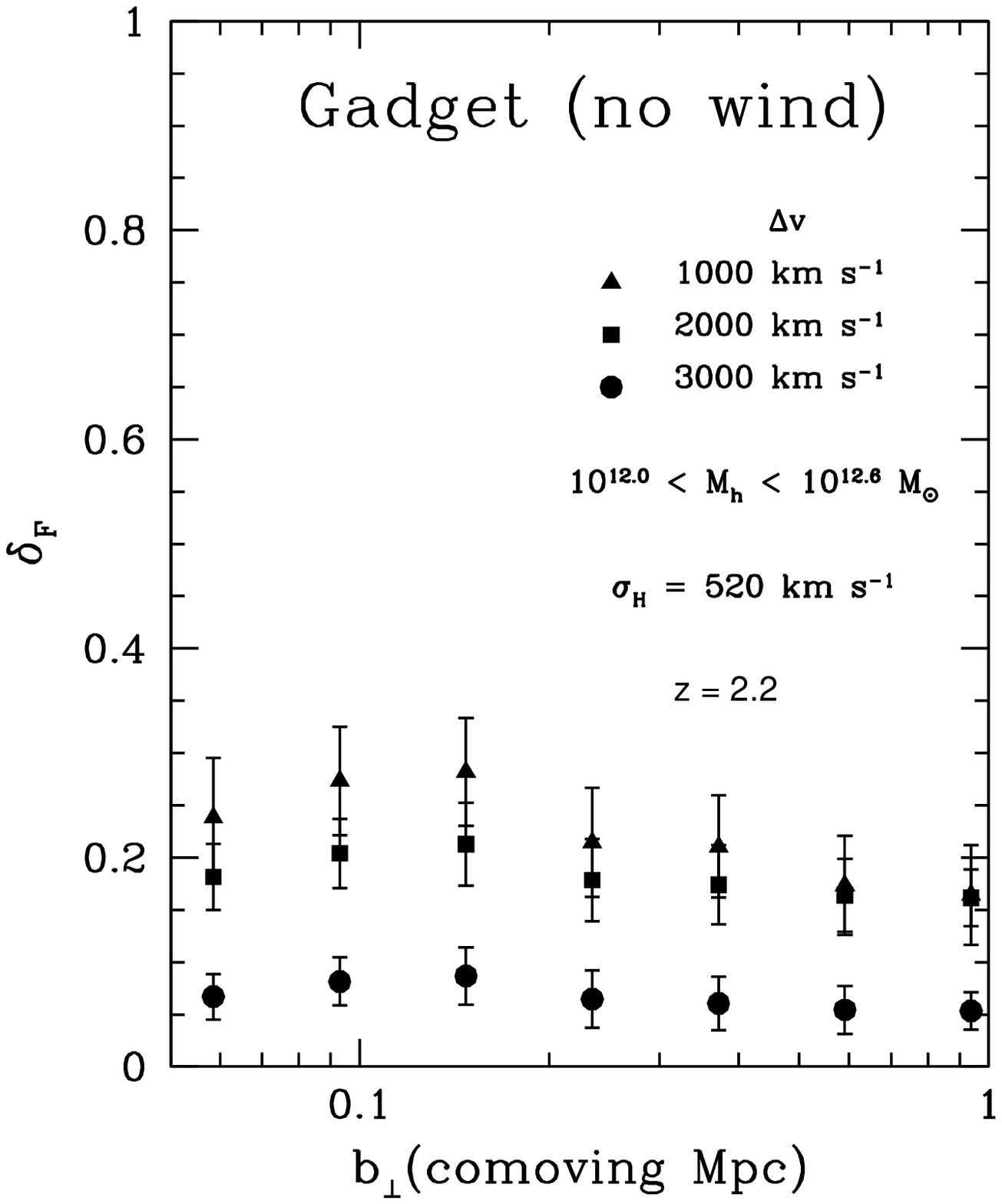}}
\vspace{-1.5cm}
\caption{Fractional absorption excess $\delta_F$ relative to the
  mean IGM for halo masses $12.0<\log_{10}M_h<12.6$ within velocity
  windows of width $\Delta v=1000\kms$ (triangles), $2000\kms$
  (squares) and $3000\kms$ (circles).  The results are centred on the
  halo centre-of-mass velocity as a function of the line-of-sight
  impact parameter $b_\perp$ for the \texttt{GADGET-3} non-wind
  simulations (G30sfnw) at $z=2.2$. While increasing the velocity
  window reduces the variance, it also suppresses the signal. The
  window $\Delta v=2000\kms$ is a good compromise. Halo velocities
  include a random component drawn from a Gaussian distribution with
  standard deviation $\sigma_{\rm H}=520\kms$.  }
\label{fig:df_conv_gadget_velw_P13}
\end{figure}

Increasing the velocity window width $\Delta v$ suppresses the values
of $\delta_F$ at all impact parameters within the turnaround radius of
massive haloes, while reducing the spread in values, as shown in
Fig.~\ref{fig:df_conv_gadget_velw_P13}. Decreasing $\Delta v$ from
$2000\kms$ to $1000\kms$ slightly increases the signal while producing
a somewhat wider spread. A window width of $\Delta v=2000\kms$
produces a good compromise between these competing effects. Velocity
window widths $\Delta v=1000\kms$ and $2000\kms$ to compute $\delta_F$
are used by \citet{2012ApJ...751...94R} and
\citet{2013ApJ...776..136P}, respectively, much wider than the
respective typical halo velocity errors of $\sigma_H=130\kms$ and
$520\kms$. We have confirmed that a halo velocity uncertainty as large
as half the velocity window affects the values of $\delta_F$ by less
than two per cent.

\begin{figure}
\scalebox{0.5}{\includegraphics{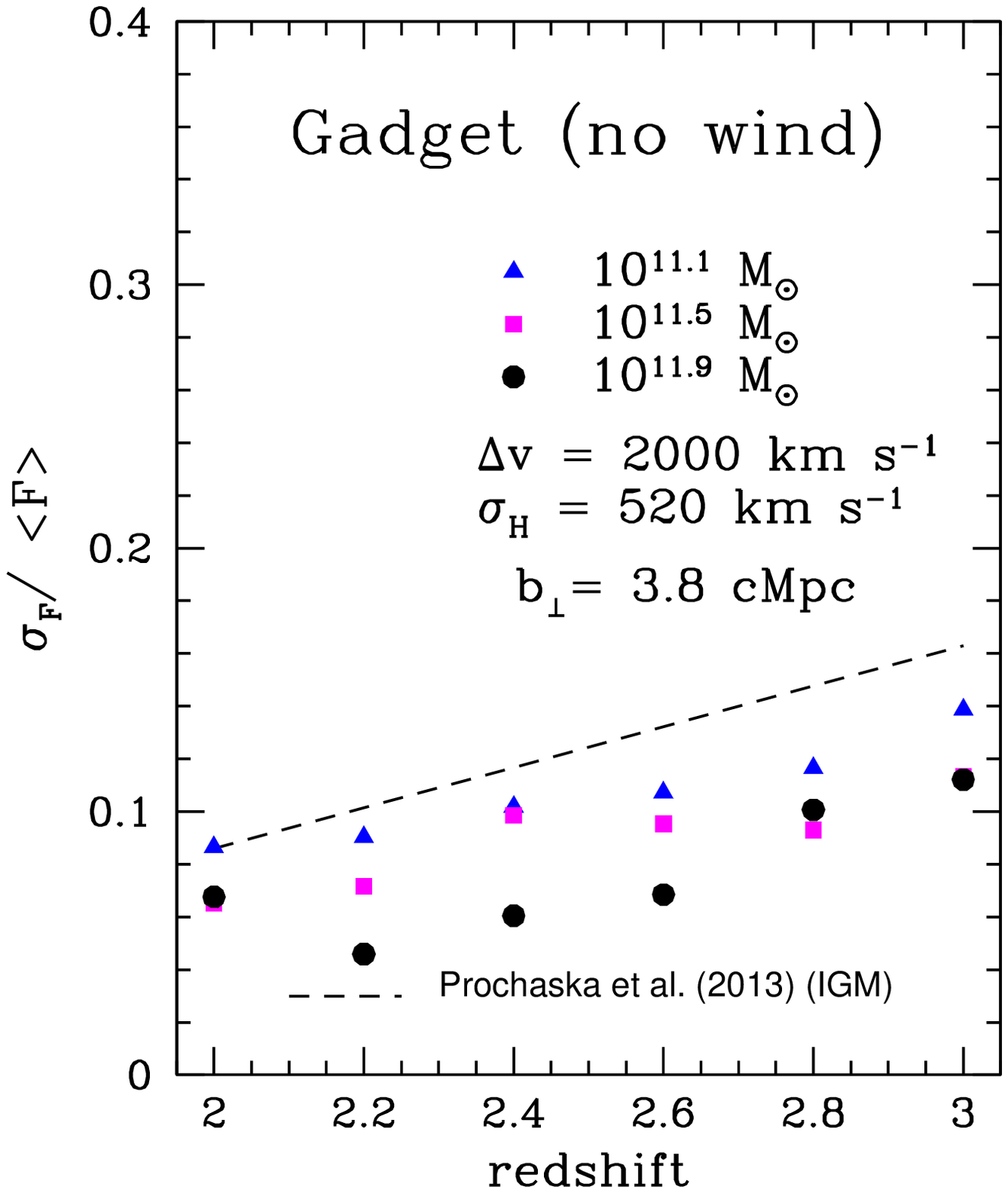}}
\vspace{-1.5cm}
\caption{Relative flux fluctuation $\sigma_F/\langle F\rangle$
  averaged over a velocity window $\Delta v=2000\kms$ centred on the
  halo centre-of-mass velocity for halo masses $\log_{10}M_h=11.1$
  (triangles; blue), 11.5 (squares; magenta) and 11.9 (circles; black)
  at $b_\perp=3.8$~Mpc (comoving).  The results are displayed as a
  function of redshift for the \texttt{GADGET-3} non-wind simulation
  (G30sfnw). The dashed line is the fit from Prochaska et al. (2013)
  to IGM measurements. It is most closely approached for the lower
  mass haloes. Halo velocities include a random component drawn from a
  Gaussian distribution with standard deviation $\sigma_{\rm
    H}=520\kms$.  }
\label{fig:sigF_evol_Gadget_nw}
\end{figure}

Finally, the relative fluctuations in the Ly$\alpha$ flux in a given
velocity window in random diffuse IGM regions increase with
redshift. In Fig.~\ref{fig:sigF_evol_Gadget_nw}, we show the
convergence in our simulations to the IGM data measured in a spectral
window of $\Delta v=2000\kms$ \citep{2013ApJ...776..136P} for large
impact parameters around the haloes. The diffuse IGM value is
approached most rapidly in the gaseous surroundings of the lower mass
haloes, for which the mesogalactic region is smaller.

\label{lastpage}

\end{document}